\documentclass[11pt,a4paper]{article}
\usepackage[english]{babel}
\usepackage{siunitx} 
\usepackage{graphicx}
\usepackage{mathrsfs}
\usepackage{authblk}
\usepackage{hyperref}
\usepackage{amsmath}
\DeclareMathOperator{\Tr}{Tr}
\usepackage{amssymb}
\usepackage{hyperref}
\usepackage{appendix}
\usepackage{IEEEtrantools}

\title{Analysis of scattered higher dimensional data using generalized
  Fourier interpolation}

\author[a]{K.~Gellerstedt \thanks{\href{mailto:karl.gellerstedt@fysik.su.se}
    {karl.gellerstedt@fysik.su.se}}}
\author[a]{J.~Sj\"olin\thanks{\href{mailto:sjolin@fysik.su.se}
    {sjolin@fysik.su.se}}}
\affil[a]{Stockholm University, Department of Physics}

\begin{document}

\begin{titlepage}

\maketitle

\begin{abstract}
  A method based on orthogonal function series interpolation
  of the square root probability density to analyze higher dimensional
  scattered data is
  presented. The method is targeted for the use-case when
  the model and/or data are available only as discrete events.
  While fast and efficient algorithms are well known for pseudo-spectral
  (grid node based) methods, this work focuses on a spectral
  (non grid based) approach.
  A typical application is the extraction of physics model parameters
  from events detected
  in high energy particle collisions. Several examples are provided and
  the performance is compared to existing conventional procedures.
  In some cases the method can be shown to behave as an
  optimal observable of the data, exemplified by the ability to approach
  the Cramer-Rao bound.
\end{abstract}

\end{titlepage}

\clearpage

\section{Introduction}

An important step in data analysis is to estimate the probability
density\footnote{Throughout this text the term \emph{density} will be
  considered synonymous with \emph{probability density}.} of the model and
the data. In many cases, e.g. high energy collider physics, the model is of
such complexity that it can only be made available as Monte Carlo
simulated events\footnote{The term \emph{event} denotes a unit of a statistical sample.} and the complete density can only be interpolated from the
events. In
low dimensions a fast and effective way to estimate the density is to create
a histogram of the events. Another more analytical approach is to approximate
the density with an orthogonal function series (generalized Fourier series),
for an overview see e.g. Ref.~\cite{OSDE}. Unfortunately, both these methods
encounter problems as soon as the dimensionality increases.

Any attempt to perform analysis in high $d$-dimensional spaces ($d \geq O(10)$)
must be able
to cope with the so called ``curse of dimensionality''. One way to
see this is to construct a $d$-dimensional space as a tensor product
of $1$-dimensional spaces of orthogonal functions $\phi_k(x)$ of degree $k$.
For two dimensions, labeled $x$ and $y$, with eigenfunctions up to second
degree, the full tensor becomes
\begin{align*}
  f(x,y)&=a_0\phi_0(x)\phi_0(y)+a_1\phi_1(x)\phi_0(y)+a_2\phi_0(x)\phi_1(y)+a_3\phi_1(x)\phi_1(y)+a_4\phi_2(x)\phi_0(y) \\
  &+a_5\phi_0(x)\phi_2(y)+a_6\phi_2(x)\phi_1(y)+a_7\phi_1(x)\phi_2(y)+a_8\phi_2(x)\phi_2(y).
\end{align*}
The number
of coefficients $a_u$ and eigenvectors needed will scale exponentially
as $(\max(k)+1)^d$ and quickly become intractable.
A way to postpone the exponential growth is to first realize that the terms in the series
contain products of basis functions with a total degree much larger than
practically needed.
Hence limiting the maximal order $l$ of the tensor will
postpone the disaster and lead to a scaling of the type $d+l\choose l$ which
scales as $O(d^l/l!)$ for large $d$ and fixed $l$, i.e. polynomial instead
of exponential growth with
respect to the number of dimensions, see Table \ref{tab:scaling}.
Using this \emph{monomial basis}, as an approximation of the full tensor,
is common
practice in e.g. computational economics~\cite{judd1996}, and it will be
applied throughout this work unless explicitly stated. However, an
important difference is that this work uses a spectral instead of the
common pseudo-spectral approach. This means that the data points are not
restricted to be on the grid defined by the roots of the basis function with
the highest degree.

A fundamental property of a probability density is that it is strictly
positive or zero. When approximating the density with a truncated orthogonal
function series this is no longer guaranteed. A beautiful solution to assert the
non-negative property is to require
that the density is equal to the square of a real valued Fourier amplitude.
As will be shown this requirement also has the beneficial side effect that the
measured Fourier coefficients become uncorrelated.
The square root operation required by the method is solved in the sections
below by iteratively reweighting the
Fourier series into the series representing the square root of the density.

\begin{table}
  \begin{center}
    \label{tab:scaling}
    \begin{tabular}{c|c|c|c}
      \hline
      $d$ & $l$ & Full tensor & Sparse tensor \\
      \hline
      2 & 2 & 9 & 6 \\
      2 & 10 & 121 & 66 \\
      10 & 2 & 59k & 66 \\
      10 & 10 & $2.5\cdot 10^{10}$ & 185k \\
      20 & 2 & $3.5 \cdot 10^{9}$ & 231\\
      20 & 5 & $3.6\cdot 10^{15}$ & 53130 \\
      \hline
    \end{tabular}
    \caption{Number of terms for full and monomial sparse tensors.}
  \end{center}
\end{table}

\section{Background}
\subsection{Basics}
The starting point for deriving the method is the fact that an
orthogonal function
series can be used for interpolation of sampled data.
The outlined method can in principle use any orthogonal function
series as long as the data is within its domain and the norm allows for
sampling\footnote{An example which does not work is first kind Chebychev
polynomials which have a norm with singularities at the boundaries.}. 
The selected orthogonal series should preferably be chosen to match the
properties of the approximated density to allow for a minimal required set of
eigenfunctions, e.g. spherical harmonics in case of
spherical symmetry. The derivation begins with
assuming a true one dimensional density function $p(x)$ which is defined on a
bounded intervall on the real line. This means that $p(x)$ is locally
integrable and from this also follows that $p(x) \in L^1 \cap L^2$.
Let $P(x)=np(x)$ be the scaled density where $n$ is the expected
size such that
$$
\int P(x)dx=\int np(x)dx=n.
$$
Since $P(x) \in L^2$ it means that $P(x)$
can be represented as an orthonormal real valued function series
$$
P(x)=\sum^{\infty}_{k=0} \alpha_k \phi_k(x).
$$
A natural observable for the true Fourier coefficients $\alpha_k$
given $N$ events from the density $P(x)$, is
\begin{align}\label{sfitransform}
\alpha_k&=\langle P(x), \phi_k(x) \rangle=
\int_V P(x) \phi_k(x) dx = n\int_V p(x) \phi_k(x) dx =
n\langle \phi_k(x)\rangle _{p(x)}
\simeq \frac{n}{N}\sum^{N}_{x_i  \sim p(x)} \phi_k(x_i) \nonumber \\ 
&= \{ \textrm{$N\simeq n$ when $N$ is large}\}
\simeq \sum^{N}_{x_i  \sim p(x)} \phi_k(x_i) =  a_k.
\end{align}
The expectation operator and the sum use a notation that makes the
underlying probability distribution explicit.
From the last sum it is clear that events from data or Monte Carlo
simulations which are drawn from the distribution $p(x)$ directly can be used
for an efficient projection of $p(x)$ onto the
normalized eigenvector $\phi_k(x)$. The extension to weighted events is straight forward, just substitute $\phi_k(x_i) \to \phi_k(x_i)w_i$.

\subsection{Filtering}
The observable for the coefficients $\alpha_k$ built from sampled projections in
Eq.~(\ref{sfitransform}) can be turned into an optimal observable for $P(x)$
by the use of optimal filtering.
An optimally filtered orthogonal series\footnote{Throughout this text, $\mathbf{a}$ will denote a vector or vector valued function, $a_i$ will denote the i:th element of $\mathbf{a}$ while $\mathbf{a}_n$ is the n:th instance of $\mathbf{a}$.}
$$
P(x,\lambda(N), \mathbf{a})=\sum^\infty_k \lambda_k(N) a_k \phi_k(x),
$$
can be found \cite{mise} by minimizing the mean integrated squared
error (MISE)
$$
\min_{\lambda(N)} \int_V |P(x)-P(x,\lambda(N),\mathbf{a})|^2dx.
$$
The solution for the optimal filter coefficients is
$$
\lambda_k(N)=\frac{\alpha^2_k}{\alpha^2_k+N\textrm{var}(\phi_k(x))}.
$$
Unfortunately in practice these coefficients are not easily obtained,
since both $p(x)$ and its true coefficients $\alpha_k$ are unknown.
Instead for large $N$ a much more useful near
optimal solution is given by $\lambda_k(N)=1$ for $k \leq M(N)$ and
$\lambda_k(N)=0$ for $k > M(N)$, where $M(N)$ is a tuned highest order
truncation. This means that $P(x,\lambda(N), a)$ is replaced by
$$
P(x,M,\mathbf{a})=\sum^{M(N)}_k a_k \phi_k(x),
$$
The extension to higher dimensions is done by forming a tensor product
of one dimensional orthogonal series for each dimension.

\section{The SFI method}
The construction of the generalized Fourier series in higher dimensions
outlined below will be referred to as Sparse Fourier Interpolation
(SFI). The key components of SFI are the sparse monomial
basis, to postpone the curse of dimensionality, and a squared Fourier amplitude
mapping asserting a strictly positive approximation. The function $P(x)$
is assumed to be approximated by a tensor build from products of
truncated one-dimensional orthonormal eigenfunctions $\phi_k(x)$ of degree $k$.

\subsection{Basis construction}
Two orthonormal sets have been found to be useful for general purposes:
$$\phi_k(x)=\sqrt{\frac{2k+1}{2}}P_k(x)$$
where $P_k(x)$ are the Legendre functions on the interval -1 to 1, and
\begin{align}
  \phi_k(x)&=\sqrt{2}\cos(k\pi x), k>0,\\
  \phi_0(x)&=1,  \nonumber
\end{align}
which are the orthonormal
finite Fourier cosine functions on the interval 0 to 1\footnote{In the code implementation the interval $[0,1]$ is mapped to $[-1,1]$ to simplify the
use of the Legendre eigenfunctions.}. In the
following examples the finite cosine functions are mainly used since they in these cases seem to work better. The density is expressed in the Fourier
coefficient vector $\mathbf{a}$ as:
$$
P(\mathbf{x},\mathbf{a}) \simeq \sum^{M(N)}_{u} a_{u} \phi_j(x_1)\phi_k(x_2)...
=\sum^{M(N)}_{u} a_{u} \phi_u(\textbf{x}),
$$
where $M(N)$ is the number of eigenvectors when $j+k+...<=l$, and $l$ is
the highest sum of degrees present in $P(\mathbf{x},\mathbf{a})$.

\subsection{A positive probability density -- the square root and diagonalization}

Even if $P(\mathbf{x})$ is originally non-negative
, there is no such guarantee for $P(\mathbf{x},\mathbf{a})$
after the truncation of the Fourier series. To assert a non-negative
density, the method assumes that $P(\mathbf{x})$ can be written as
$$P(\mathbf{x}) = A^2(\mathbf{x}) \simeq A^2(\mathbf{x},\mathbf{b})$$
and the problem is recasted as finding the coefficients in the Fourier series $A(\mathbf{x},\mathbf{b})$, and $A$ is not required to be positive. To differentiate between the previous linear SFI transform we refer to the method of finding $A$ as the square root transform, or simply SFI. Since $A(\mathbf{x},\mathbf{b})$ appears squared, Plancherel's theorem can be applied to $P(\mathbf{x})$
$$
n = \int_V P(\mathbf{x}) d\textbf{x} \simeq \int_V A^2(\mathbf{x},\mathbf{b}) d\textbf{x}=\sum_u |b_u|^2.
$$
It turns out that the noise modelling in this case is greatly simplified since the covariance
matrix of $\mathbf{b}$ becomes diagonal.

Assume the coefficients $\mathbf{b}$ are estimated using an unbinned extended maximum likelihood (EML)~\cite{emle}, with likelihood function
\begin{align}\label{eq:unb}
l(\mathbf{b}) = \ln(L(\mathbf{b}))=\sum_{\mathbf{x}_i \sim p(\mathbf{x})} \ln(A^2(\mathbf{x}_i,\mathbf{b}))-n(\mathbf{b}).
\end{align}

By taking the second
derivative of Equation \eqref{eq:unb} one can show
that for the square root transform the coefficient covariance matrix converges
towards a diagonal matrix with $1/4$ on the diagonal:
$$ V^{-1}(\hat{\mathbf{b}})_{kl} = - \frac{\partial^2\ln(L(\hat{\textbf{b}}))}{\partial b_k \partial b_l} 
= \sum_{\mathbf{x}_i \sim p(\mathbf{x})} 2 \frac{\phi_k(\textbf{x}_i)\phi_l(\textbf{x}_i)}{A^2(\textbf{x}_i, \hat{\textbf{b}})} + 2\delta_{kl} 
\simeq 2 \int_V P(\mathbf{x}) \frac{\phi_k(\textbf{x})\phi_l(\textbf{x})}{A^2(\textbf{x}, \hat{\textbf{b}})} d\textbf{x} + 2\delta_{kl}
\simeq 4 \delta_{kl}$$ 
where $\hat{\mathbf{b}}$ is the EML estimate of the coefficients and the orthonormality of $\phi_k(\textbf{x})$ as well as the assumption that $A^2(\textbf{x}, \textbf{b})$ point-wise approximates $P(\textbf{x})$ have been used.

\subsection{The square root SFI transform}

The square root transform series $A(\mathbf{x},\mathbf{b})$ can be found
by applying event
reweighting. This unfortunately requires knowledge of the function
$A(\mathbf{x},\mathbf{b})$ itself, which originally is unknown since that is
the very function to be solved for in the first place.
This recurrence problem can be circumvented assuming a start vector for the
function and then finding the solution as the fix-point from successive
iterations. 
The task in each iteration is to find $\mathbf{b}^{(n+1)}$ in $A(\mathbf{x},\mathbf{b})$, given events of $P(\mathbf{x})$ and the previous suggestion $\mathbf{b}^{(n)}$. Inspired by
$$ 
\int_V A(\mathbf{x})\phi_u(\mathbf{x})d\mathbf{x} = \int_V P(\mathbf{x}) \frac{\phi_u(\mathbf{x})}{A(\mathbf{x})}d\mathbf{x},
$$
the reweighting during each iteration can be written


$$
b_u^{(n+1)} = \sum_{\mathbf{x}_i \sim A(\mathbf{x})} \phi_u(\mathbf{x}_i)
\simeq \sum_{\mathbf{x}_i \sim P(\mathbf{x})}\frac{\phi_u(\mathbf{x}_i)}{A(\mathbf{x}_i, \mathbf{b}^{(n)})}
$$
where the initial guess is $b^{(0)}_0=\sqrt{N}$ and $b^{(0)}_{i\neq0}=0$.
The convergence of the iterations are controlled by a convex combination
with a relaxation parameter $\omega$ and an optional regularization parameter $\epsilon$ implemented as
\begin{align*}
\boxed{
b^{(n+1)}_u=(1-\frac{\omega}{2})b^{(n)}_u+\frac{\omega}{2} \sum^N_{\textbf{x}_i \sim p(\textbf{x})}
\frac{\phi_u(\textbf{x}_i)}{A(\textbf{x}_i,\textbf{b}^{(n)}) + \epsilon \mathrm{sign} (A(\textbf{x}_i,\textbf{b}^{(n)}))}.
}
\end{align*}
For $\omega=1$ this coincides with the Newton method for solving the
equation $P(\mathbf{b})-A^2(\mathbf{b})=0$. In the examples below
$\omega$ ranges between $0.25-1.0$.

\subsection{Minimizing the interpolation uncertainty in higher dimensions}

In higher dimensions, large amounts of training data is required to
sufficiently outnumber the degrees of freedom present in the SFI series.
For a monomial sparse series, the training dataset with $N$ events in $d$ dimensions must fulfil
$$
Nd \gg \frac{(d+l)!}{l!d!}.
$$
where $l$ is the basis function maximum degree. This is in practice often difficult to achieve.
There are several intuitive and well known techniques available to avoid too
many degrees of freedom, a.k.a. overtraining and noise learning, which are
effective also in the SFI context. The degrees of freedom can be regulated via
the \emph{meta parameters}, such as maximum degree. One powerful method to determine the values
of the meta parameters is to minimize the
\emph{cross entropy} ($H$) as a function of the meta parameter in question.
For samples drawn from the distribution $p(\textbf{x})$ the cross entropy $H$
with respect to the interpolated distribution $p(\textbf{x,\textbf{b}})$ can
be estimated with:
$$
H = -\frac{1}{N} \sum_{\textbf{x}_i \sim p(\textbf{x})} \ln(p(\textbf{x}_i,\textbf{b})).
$$
For SFI the maximum degree of the polynomial ($l$) is a direct
handle to control the degrees of freedom of the interpolation and
regulate $H$. It is strongly advised to always work with SFI transformations
that use the maximal degree determined from the optimum provided by
the cross entropy minimum. The method still works with harder truncation of
the maximum degree for the eigenfunctions, albeit with reduced performance.
However a too high maximum degree results in the usual and familiar problems
associated with overtraining.
An additional way to improve the performance is to transform the input
variables such that the required number of eigenfunctions are reduced.


\subsection{Example distributions}
\begin{figure}
  \centering
  \includegraphics[width=0.45\textwidth]{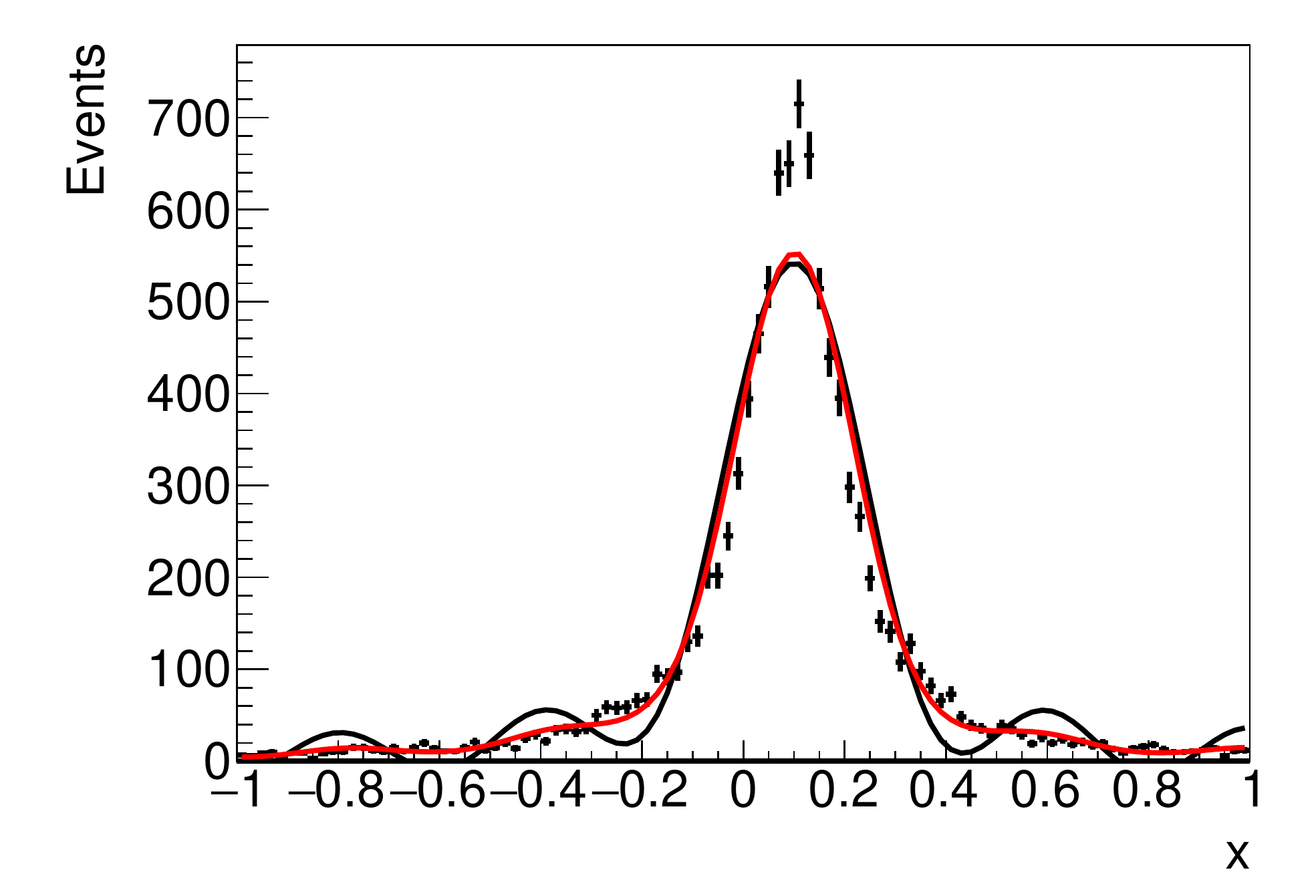}
  \includegraphics[width=0.45\textwidth]{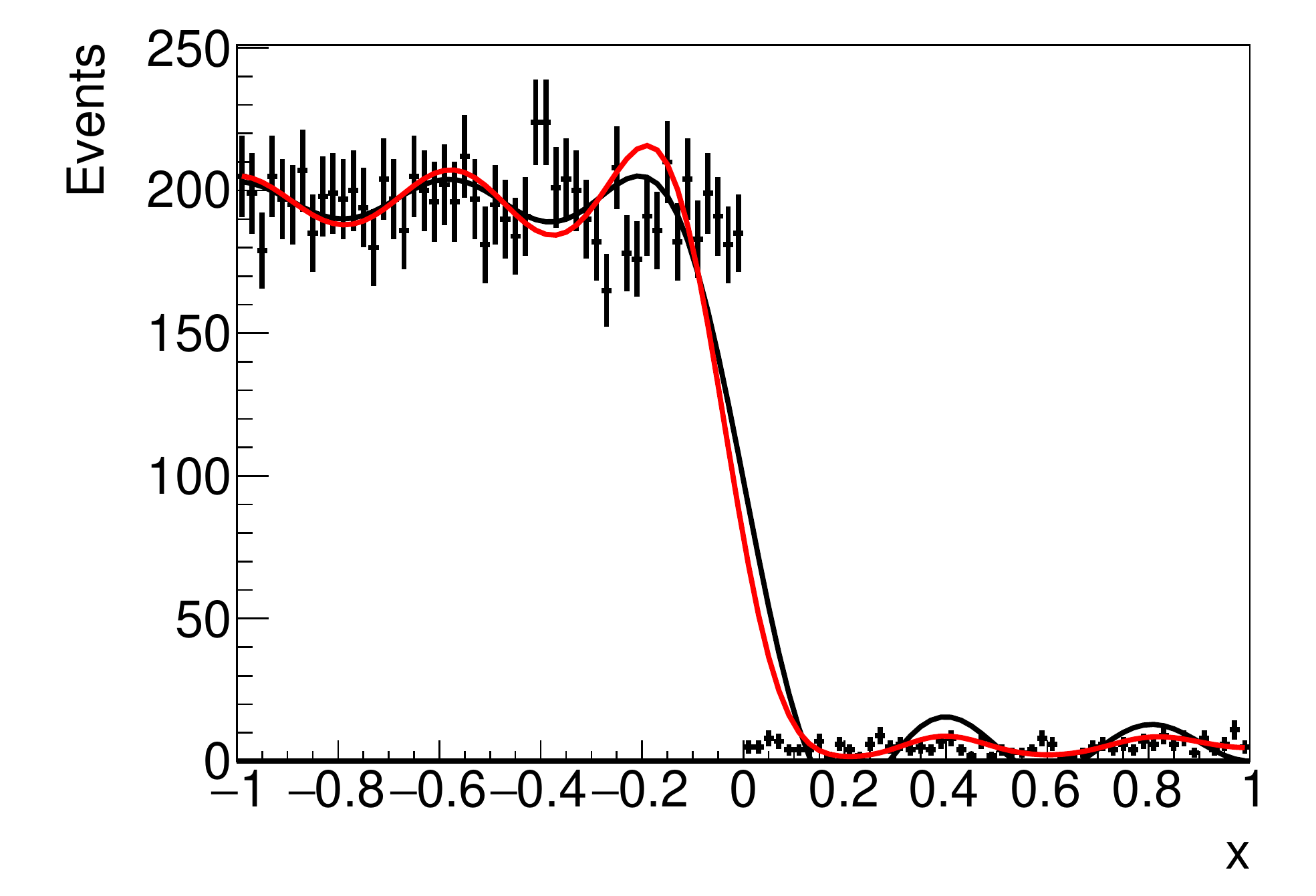}
  \caption{Histograms and SFI interpolations for Cauchy (left) and step (right) distributions with hard non-optimal Fourier coefficient truncation of the polynomial. The red curve is the square root SFI estimate. Note how the square root SFI transform remains non-negative by construction.}
  \label{fig:pdf_examples}
\end{figure}

Figure \ref{fig:pdf_examples} show histograms of Cauchy (left) and step function (right) distributed scattered data compared to SFI interpolation under hard non-optimal Fourier coefficient truncation. The red curve is the square root SFI estimate. The series use cosine as basis functions and are truncated above degree 10. Both distributions are challenging in different ways, the Cauchy (Breit-Wigner) has a sharp peak while the step function is discontinuous. In both cases it can be seen that the linear SFI transform give a negative density estimate for some fraction of the interval.

\begin{figure}
  \centering
  \includegraphics[width=0.45\textwidth]{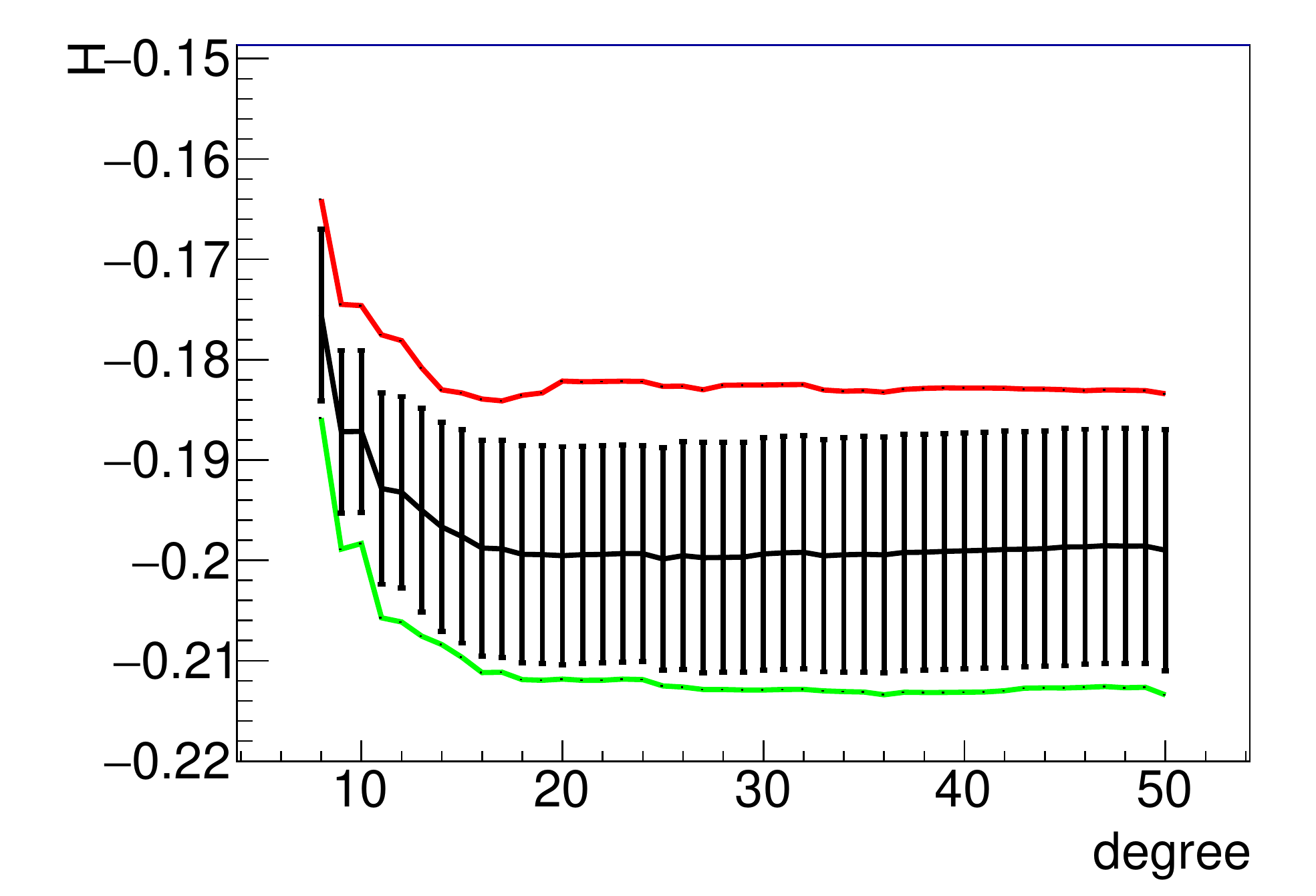}
  \includegraphics[width=0.45\textwidth]{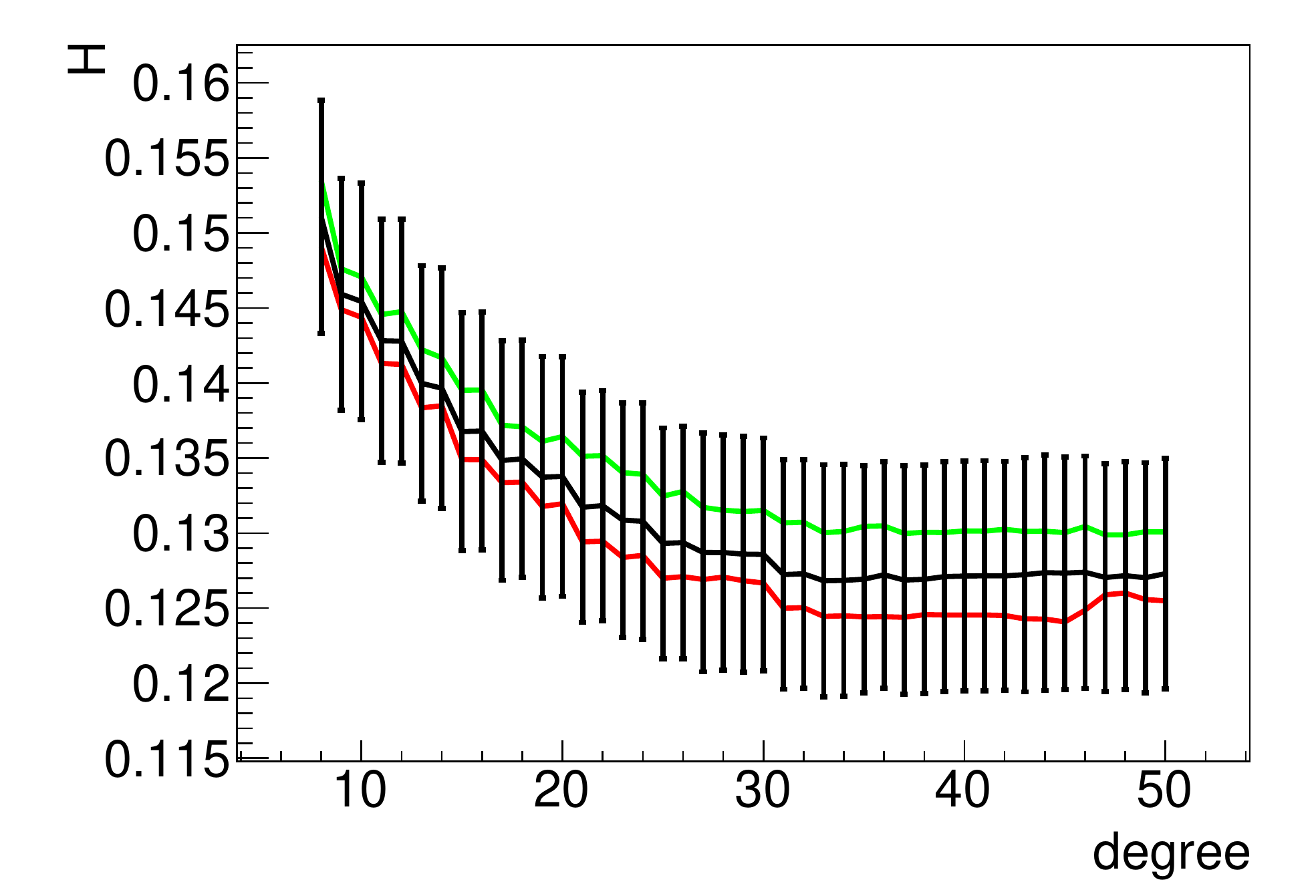}
  \caption{Estimates of the cross entropy for the Cauchy (left) and step (right) distributions. The red and green curves in the figure approximate $H$ using a transform of even samples evaluated on odd samples, and vice verse. The black curve approximates $H$ by 5-fold cross entropy where the average and uncertainty of the average is shown.}
  \label{fig:pdf_ce_examples}
\end{figure}

\begin{figure}
  \centering
  \includegraphics[width=0.45\textwidth]{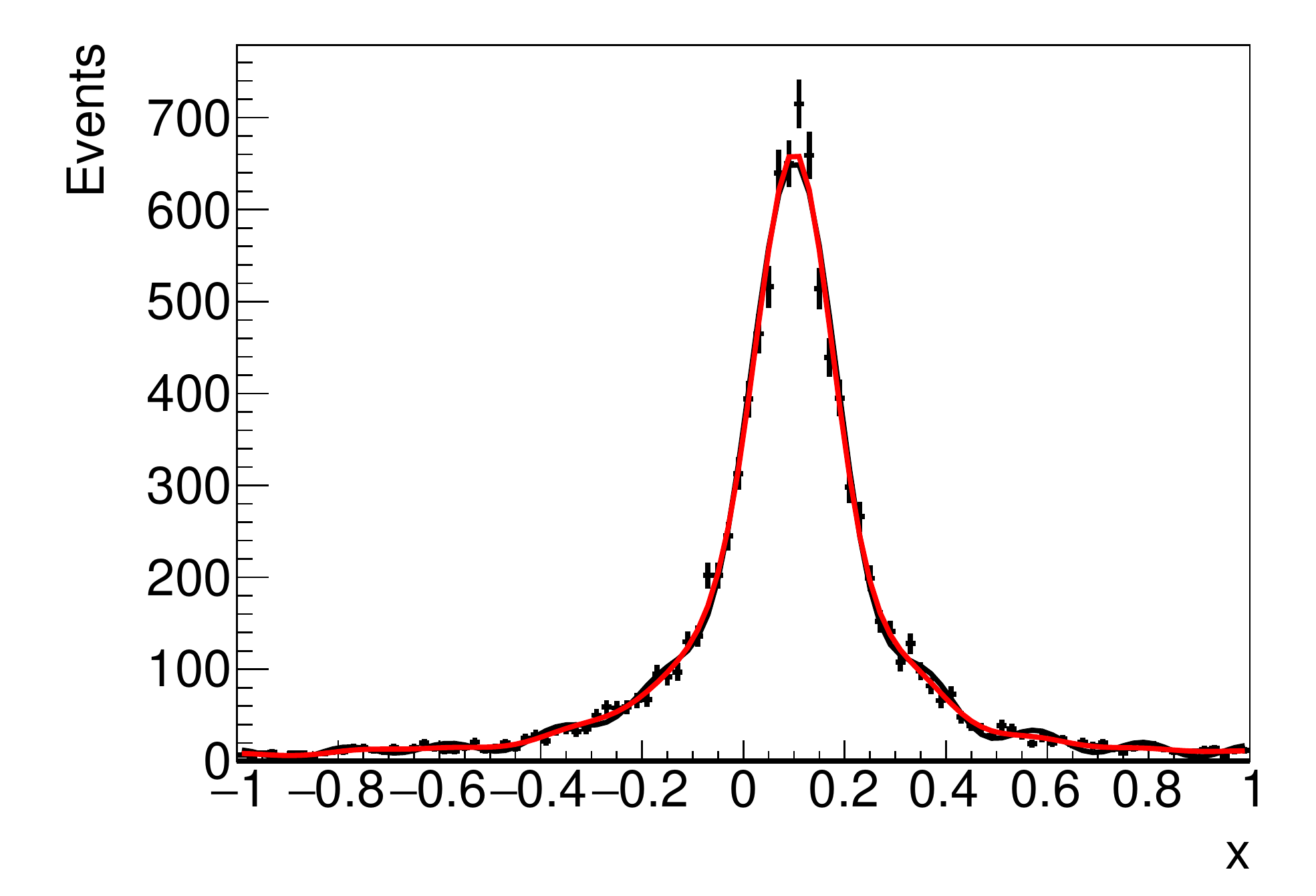}
  \includegraphics[width=0.45\textwidth]{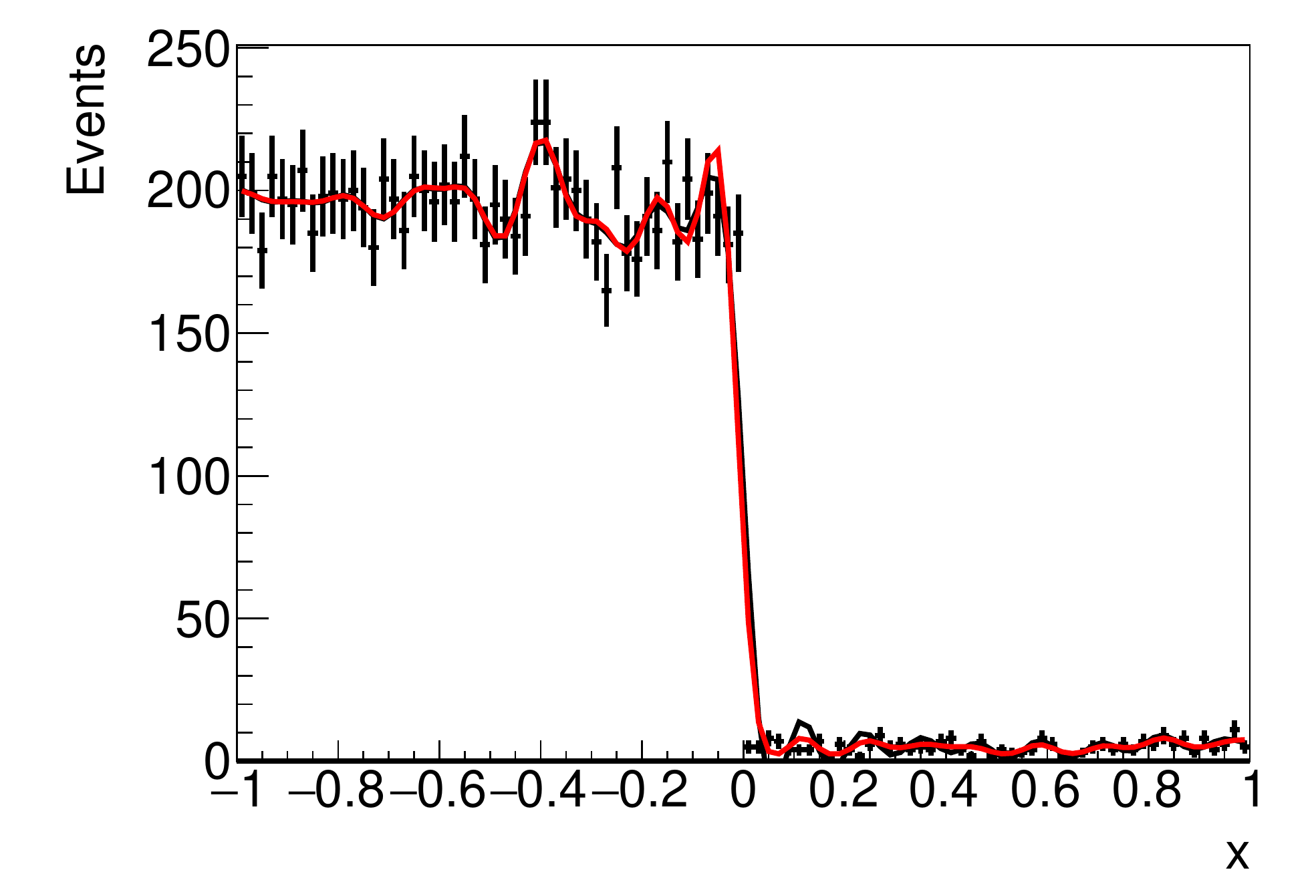}
  \caption{Histogram and optimal SFI interpolation for Cauchy (Breit-Wigner) and step distributions with maximum degree 18 and 34 respectively. The optimal degree is selected using the minimum degree of the series where the estimated cross entropy $H$ reaches its minimum.}
  \label{fig:pdf_examples_opt}
\end{figure}

Figure \ref{fig:pdf_ce_examples} show estimates of $H$ as a function of the maximum degree for the two example distributions using 10k events. The red and green curves in the figure approximate $H$ using a transform of even samples evaluated on odd samples, and vice verse. The black curve approximates $H$ by 5-fold cross entropy where the average and uncertainty of the average is shown. In the n-fold cross entropy the transform is made out of each of $n$ subsamples (defined by excluding the $i$:th event where $i$ runs from 0 to 4) evaluated on the rest of the data. The optimal polynomial degree can be estimated from the minimal degree when $H$ reaches its minimum. This gives
$l\simeq 18$ for the Chauchy example and $l\simeq 34$ for the step example. The distributions in Figure \ref{fig:pdf_examples} replaced with the estimated optimal truncation are shown in Figure \ref{fig:pdf_examples_opt}.

\section{Method validation: SFI for nonparametric density estimation}

\begin{figure}
  \centering
  \includegraphics[width=0.32\textwidth]{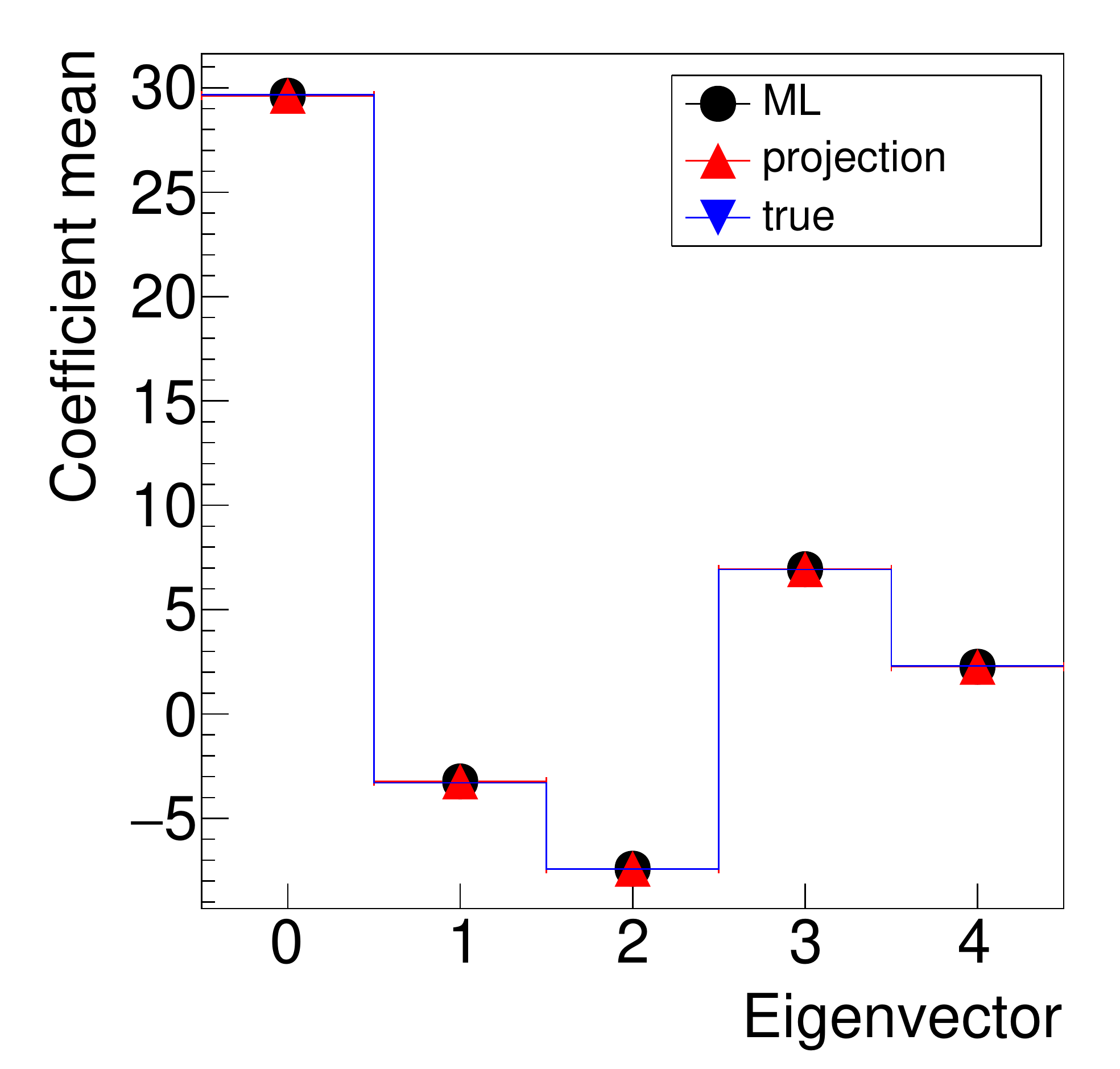}
  \includegraphics[width=0.32\textwidth]{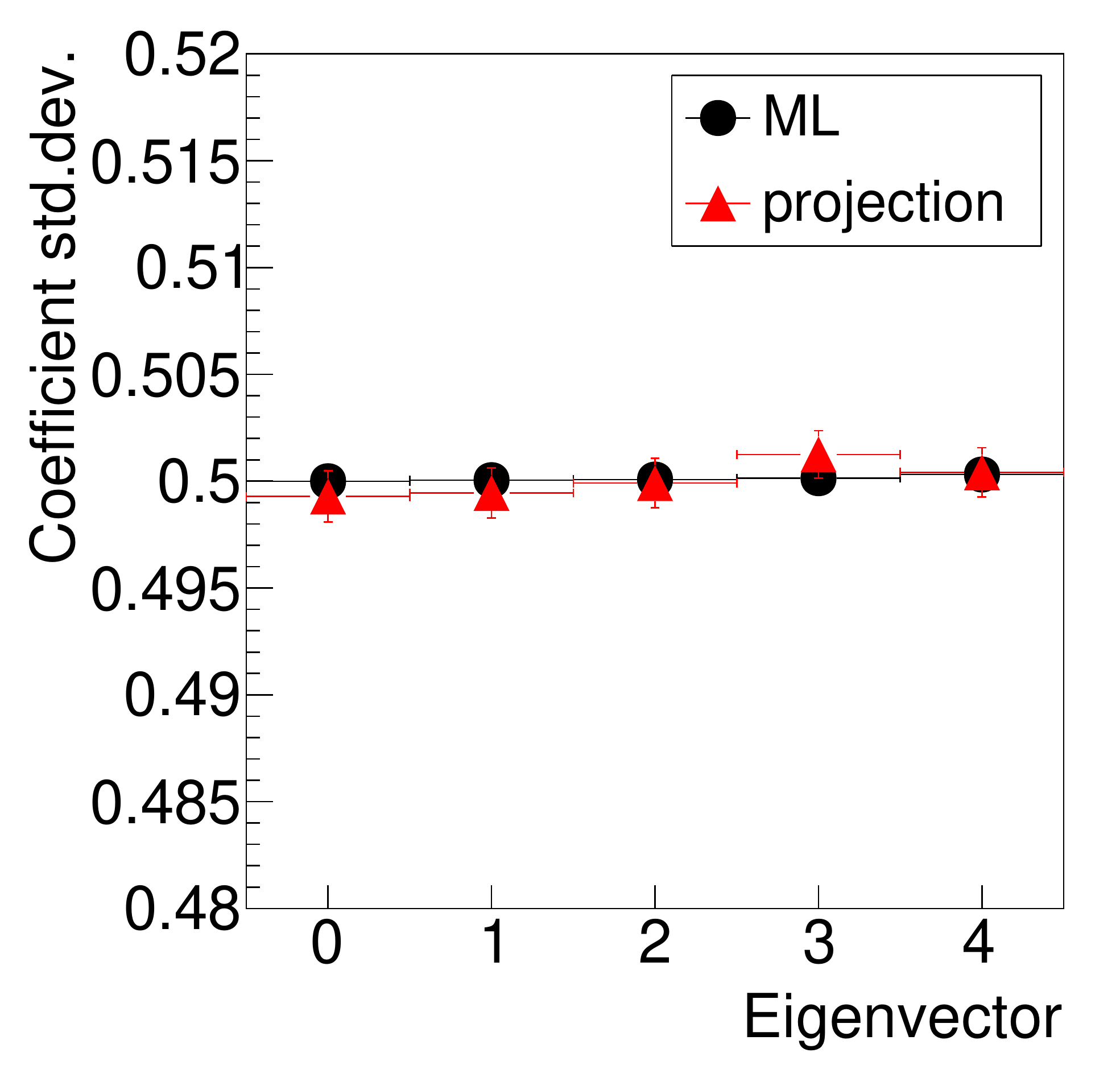}
  \includegraphics[width=0.32\textwidth]{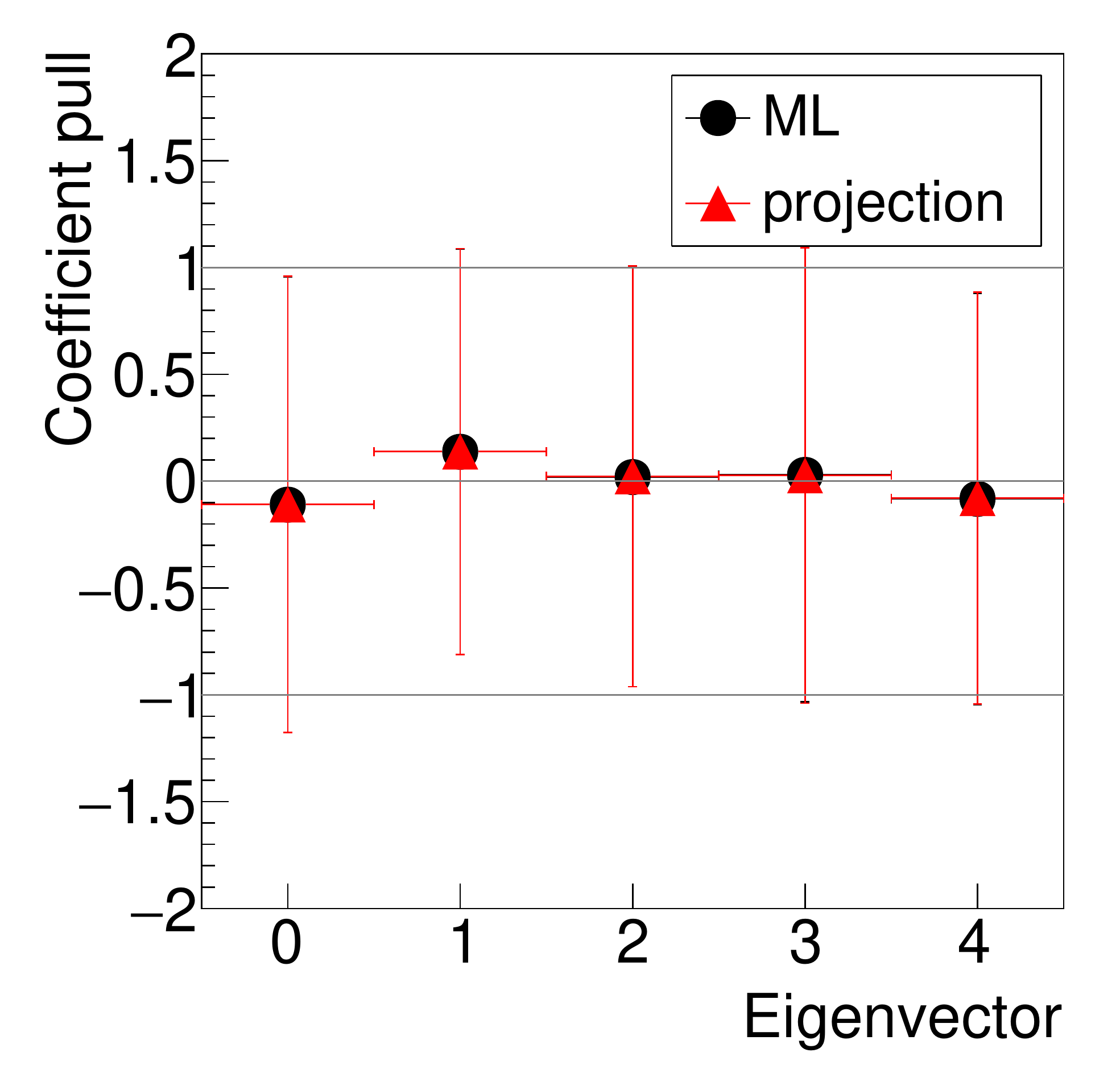}
  \caption{Comparing the performance of an extended maximal
    likelihood fit (EML) to the direct event projection method to calculate
    the SFI transform of five eigenvectors using Poisson(10k) distributed
    events and 200 pseudo experiments.
    The plots show the mean (left), standard deviation (middle), and
    pull (right).}
  \label{fig:density_test}
\end{figure}

An optimal alternative to the SFI method described in
the previous chapter, that is only tractable in
low dimensions, is to directly estimate the Fourier coefficients $\mathbf{b}$ using an unbinned EML, with likelihood function given by Equation \eqref{eq:unb}.
The algorithm MIGRAD from the package MINUIT~\cite{minuit} is used for
the numerical minimizations.
The density is estimated using a sample of Poisson(10k) distributed events
generated from a density made out of five
known orthonormal cosine eigenvectors. The performance of using either EML or
SFI transformations are compared in Figure \ref{fig:density_test}.
The uncertainties for SFI are estimated using bootstrapping.
The coefficients estimated with SFI are found to be statistically compatible
with EML and
consequently compatible with the Cramer-Rao bound.

That the covariance matrix becomes diagonal is confirmed by the
estimated covariance matrix provided by the EML fit, and is in stark contrast to
the linear transform where the Fourier coefficients are highly correlated
and in general requires the full covariance matrix to be useful. 
However, as will be shown later, the covariance matrix for the linear
transform can be computed from the coefficients.

\begin{figure}[!hbp]
  \centering
  \includegraphics[width=0.9\textwidth]{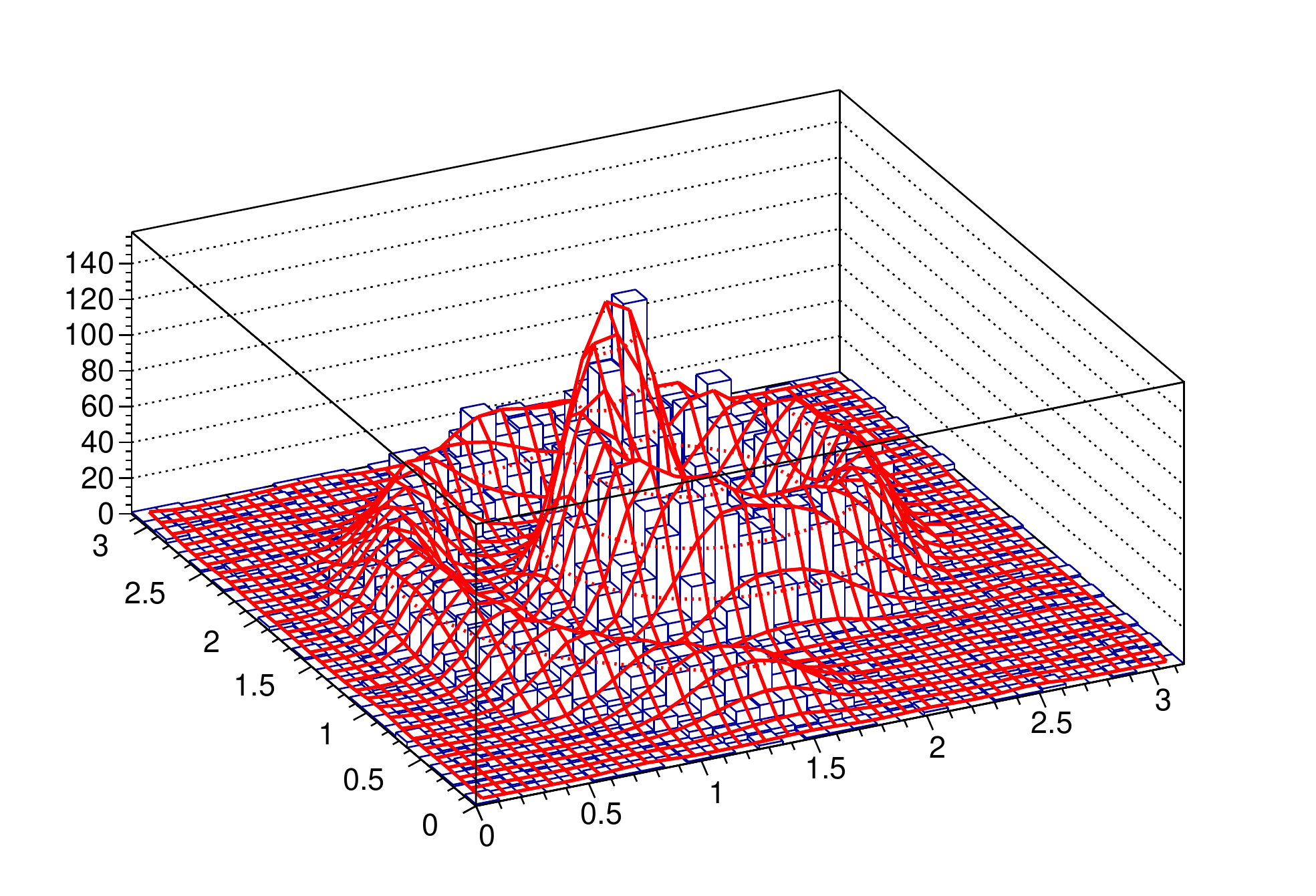}
  \caption{The spiral density estimated using SFI interpolation with $l=15$ and 136
    cosine eigenvectors.}
  \label{fig:spiral}
\end{figure}
The density of a more challenging 2D spiral
distribution is shown in Figure \ref{fig:spiral}. The density estimation
is performed using a series with $l=15$ giving 136
eigenfunctions, both using EML and SFI transform with compatible results.
Given a computing time of less than a minute on a standard laptop the
EML fit is practically limited to $O(100)$ eigenvectors,
while the SFI transform limitation in the current implementation
is $O(100k)$ eigenvectors. Using the numbers provided in
Table~\ref{tab:scaling} shows that this allows for access to domains
that are normally out of reach, e.g. $d=10$ and $l=10$.

\section{SFI example: classification}
Given signal and background densities it is possible to perform
classification. This is not the primary use-case for SFI but rather a way
of comparing the performance of SFI to other well established techniques.
The signal decision function is constructed from separate SFI
transforms of the signal and background model events according to
$$
s(\mathbf{x})=\frac{n_sp_s(\mathbf{x})}{n_sp_s(\mathbf{x})+n_bp_b(\mathbf{x})}
=\frac{P_s(\mathbf{x})}{P_s(\mathbf{x})+P_b(\mathbf{x})}
\simeq \frac{P_s(\mathbf{x}, \mathbf{a}_s)}{P_s(\mathbf{x}, \mathbf{a}_s)+P_b(\mathbf{x}, \mathbf{a}_b)}.
$$

\begin{figure}[!hbp]
  \centering
  \includegraphics[width=0.45\textwidth]{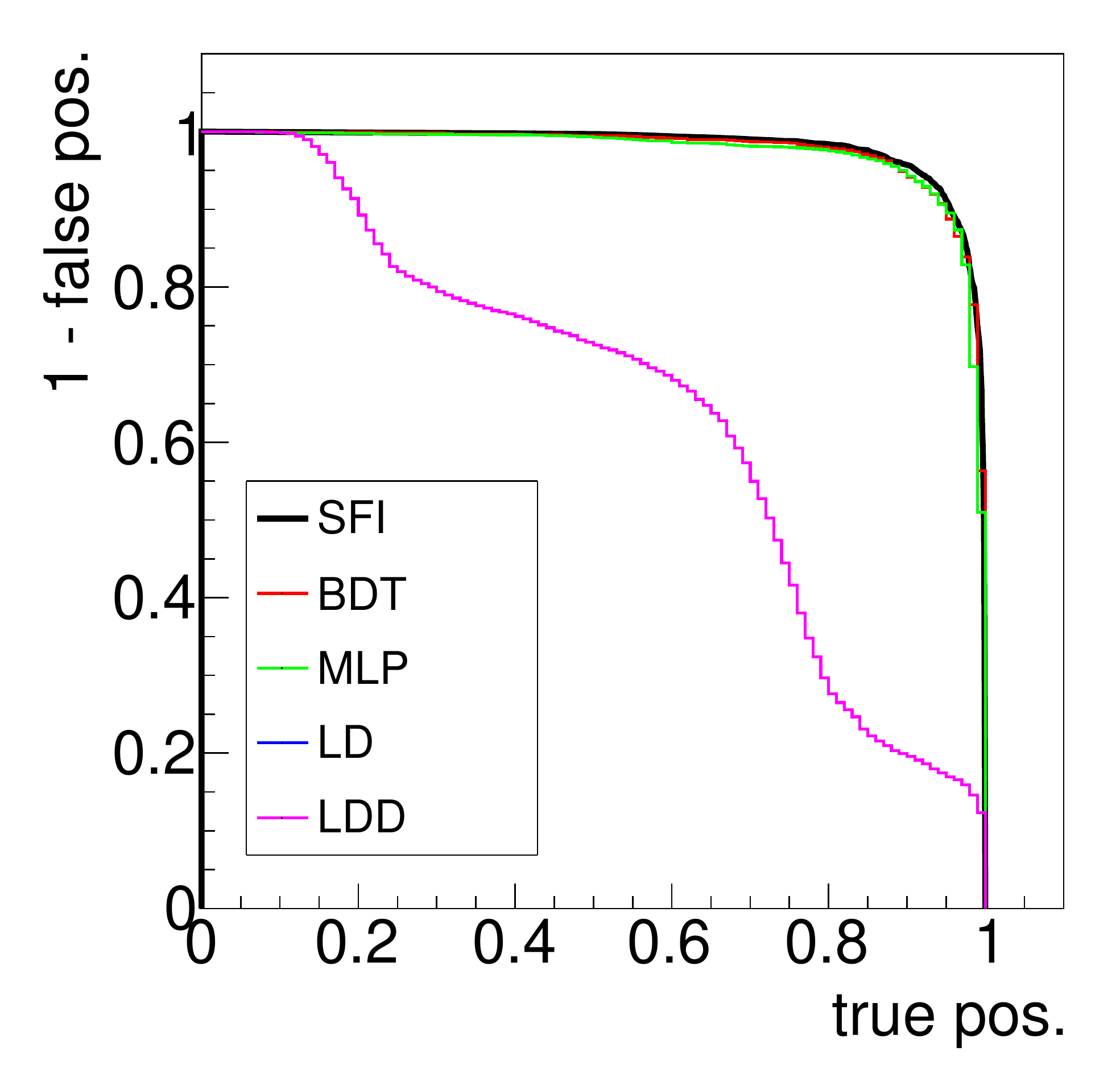}
  \includegraphics[width=0.45\textwidth]{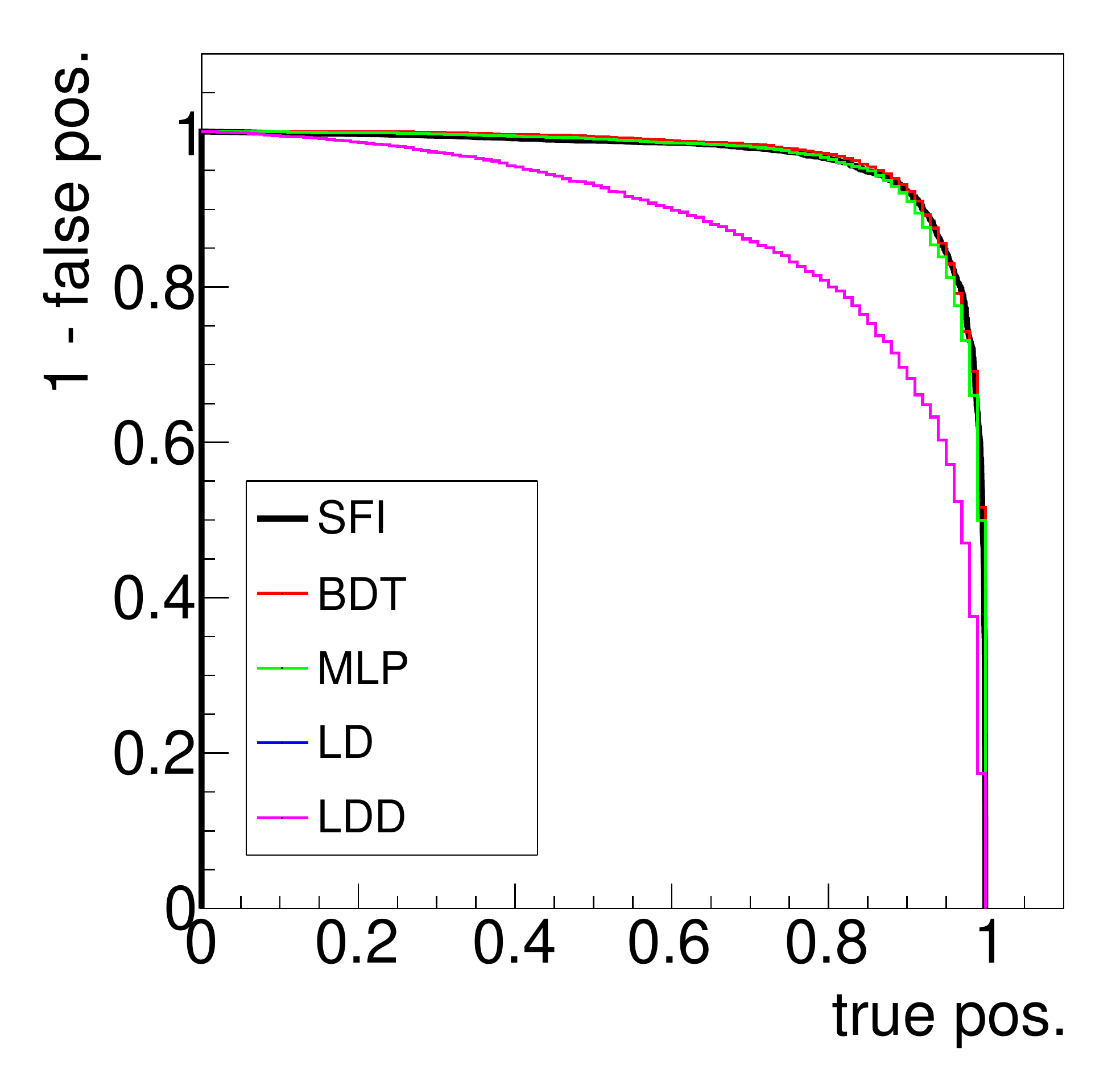}
  \caption{ROC curve for the spiral (left) and HEP (right) classification
    comparing SFI to other standard
    classifications methods: neural networks (MLP), boosted decision trees
    (BDT), naive Bayes classifier (LD) or de-correlated LD (LDD).}
  \label{fig:ROC_spiral}
\end{figure}

In these examples $n_s = n_b$. The first example uses intertwined spirals of the type shown in the previous
chapter, but in this case with $l=12$ to limit overtraining. 
The classification boundary is highly nonlinear and cannot be well
separated using only a naive Bayes classifier (LD) even if the input
variables are de-correlated (LDD). The SFI transform is compared to
neural networks (MLP), boosted decision trees (BDT) and LD(D) with respect
to timing and classification performance. The alternative classifiers are
evaluated with the TMVA framework \cite{tmva}. Classification performance is
measured as received operation characteristics (ROC) and area under
ROC curve (AUC) in the test sub sample, see
Table \ref{tab:AUC_spiral} and Figure \ref{fig:ROC_spiral}. 
A comparison between the classification efficiencies for the different
methods can be found in Appendix \ref{class_cmp}. 
More details on the example are given in Appendix \ref{apx:spiral_details}.

\begin{figure}[!hbp]
  \centering
  \includegraphics[width=0.45\textwidth]{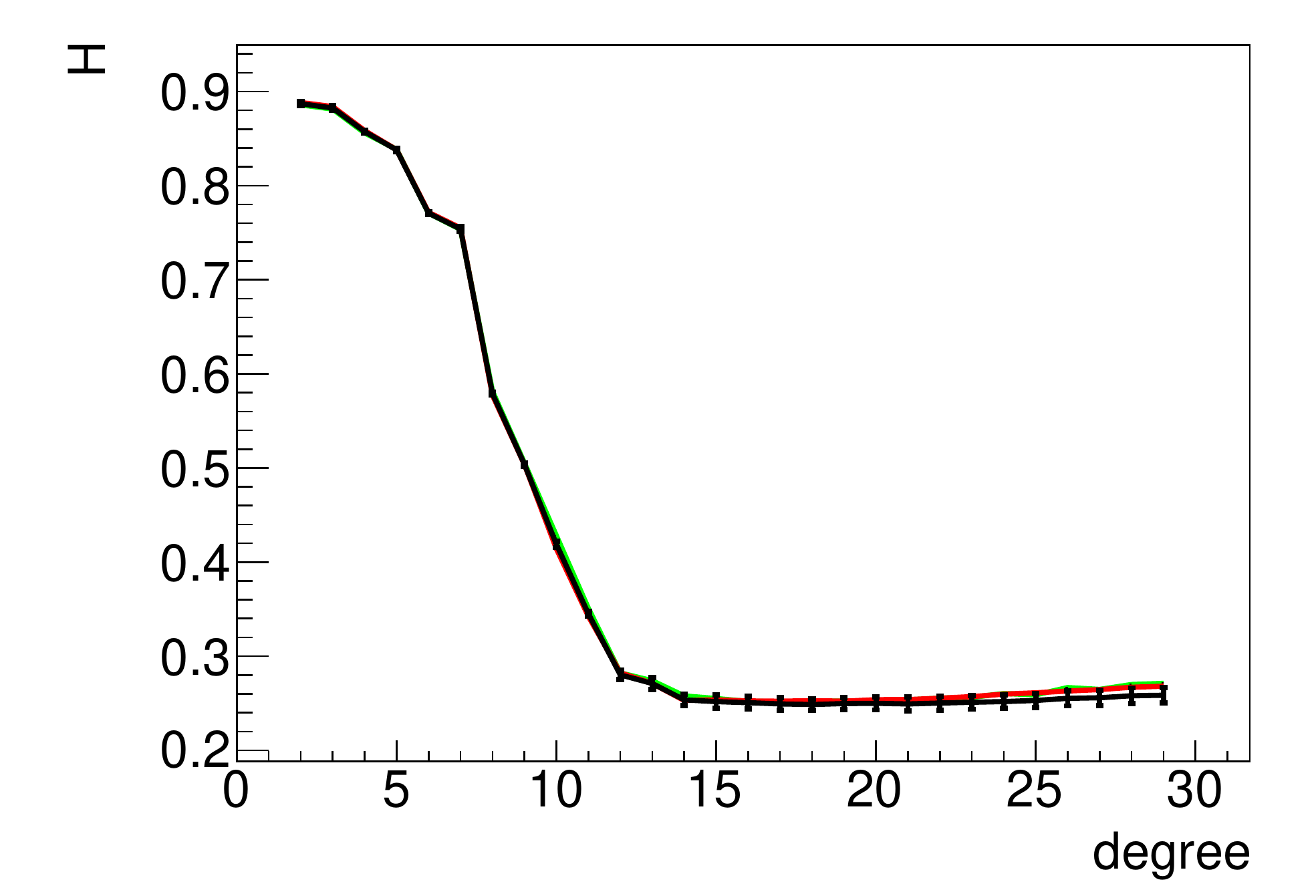}
  \caption{Estimates of the cross entropy for the the spiral density.}
  \label{fig:spiral_ce}
\end{figure}

The choice of $l=12$ is a trade off between speed and accuracy. The cross entropy for this density as function of $l$ is shown in Figure \ref{fig:spiral_ce}. As can be seen the choice of $l=12$ is below the region where over training begins. 

A second classification example is taken from high energy physics (HEP) to
validate SFI in higher dimensions. The classification task is to separate
top quark pairs from Z bosons in the dilepton plus at least one jet final state.
The density has six dimensions: transverse momentum and $\eta$ of
the two leptons, invariant mass of the two leptons, and transverse momentum
of the leading jet. The input variables are transformed for all methods, 
but for SFI the variables for $P_s(\mathbf{x}, \mathbf{a}_s)$ and $P_b(\mathbf{x}, \mathbf{a}_b)$ are transformed differently, 
since these densities are represented as different series and the extra variable transform can be applied
with little additional cost.
Classification performance is measured as ROC and AUC, see
Table \ref{tab:AUC_hep} and Figure \ref{fig:ROC_spiral}. 
A comparison between the classification efficiencies for the different
methods can be found in Appendix \ref{class_cmp}.
More details on the example are given in Appendix \ref{apx:hep_details}.

\begin{table}[!hbp]
  \center
\begin{tabular}{l  r @{.} l  r @{.} l  r @{.} l }
\hline
  & \multicolumn{2}{c}{Training time (s)}  & \multicolumn{2}{c}{Test time (s)}  & \multicolumn{2}{c}{AUC Test} \\
\hline
SFI 
 & 0&10  & 0&01  & 0&98 \\
MLP 
 & 7&93  & 0&01  & 0&97 \\
BDT 
 & 0&60  & 0&09  & 0&98 \\
LD 
 & 0&00  & 0&00  & 0&65 \\
LDD 
 & 0&01  & 0&01  & 0&65 \\
\hline
\end{tabular}
\caption{Spiral classification training time (for 10k events) and area under curve (AUC)
  numbers for different classifier methods\label{tab:AUC_spiral}.}
\end{table}

\begin{table}[!hbp]
  \center
\begin{tabular}{l  r @{.} l  r @{.} l  r @{.} l }
\hline
  & \multicolumn{2}{c}{Training time (s)}  & \multicolumn{2}{c}{Test time (s)}  & \multicolumn{2}{c}{AUC Test} \\
\hline
SFI 
 & 0&09  & 0&04  & 0&97 \\
MLP 
 & 8&86  & 0&01  & 0&97 \\
BDT 
 & 0&27  & 0&04  & 0&97 \\
LD 
 & 0&01  & 0&00  & 0&88 \\
LDD 
 & 0&02  & 0&01  & 0&88 \\
\hline
\end{tabular}
\caption{HEP classification training time and area under curve (AUC)
  numbers for different classifier methods\label{tab:AUC_hep}.}
\end{table}

%

\section{SFI example: parameter estimation}

The examples below show two examples of parameter estimation. The parameter
that is estimated is the mean
of a Gaussian with a fixed standard deviation of $\sigma=0.1$ on top of
a exponentially falling background, with
Poisson distributed numbers of data events. The fraction of the true signal
and background events is a priori unknown and varied during the pull tests.
The parameter estimation examples highlights two possible distinct working
modes.
In the first (default) case the model is SFI transformed into
square root Fourier space and then inversely transformed back into the
original x-space. This transformation and inverse transformation is how
the interpolation of the original sampled model is achieved. The sampled data
points are then compared to the interpolated density using an unbinned extended
maximum likelihood.

In the second example both the model and the data are linearly SFI
transformed into Fourier space and the analysis if performed entirely
within Fourier space without any inverse transformation back to x-space.

\subsection{Default SFI}
The starting point is a transform describing the conditional pdf of the signal given the parameters $\mathbf{c}$ : $p_s(\mathbf{x}_i | \mathbf{c})$. Depending on the model this transform is constructed in one of two ways.
If model events are available uniformly covering the entire model parameter space for
$\mathbf{c}$ then $p_s(\mathbf{x}_i | \mathbf{c})$ can easily be constructed from the joint pdf $p_s(\mathbf{x}_i, \mathbf{c})$. If the model samples are given uniformly at certain discrete values of $\mathbf{c}$ then the method outlined in section \ref{sec:discrete_c} can be used.

From the conditional pdf $p_s(\mathbf{x}_i | \mathbf{c}) = p_s(\mathbf{x}_i,\mathbf{b}(\mathbf{c}))$ it is straight forward to formulate the default SFI parameter
estimation as an unbinned EML with likelihood
$$
l(n_s,n_b,\mathbf{c}) =\ln(L(n_s,n_b,\mathbf{c}))=\sum_{i=1}^{N} \ln\left(\frac{n_sp_s(\mathbf{x}_i,\mathbf{b}(\mathbf{c}))+n_bp_b(\mathbf{x}_i)}{n_s + n_b}\right)+N \ln(n_s + n_b)-n_s-n_b,
$$
where $n_s$ and $n_b$ are the model yields for the signal and background
respectively, $\mathbf{c}$ is the signal model parameter vector
with corresponding Fourier coefficients $\mathbf{b}(\mathbf{c})$, and
$\mathbf{x}_i$ are the scattered data events.
The signal and background Fourier coefficients are normalized such that the
integrals of $p_s$ and $p_b$ are equal to one.

\begin{figure}[!hbp]
  \centering
  \includegraphics[width=0.45\textwidth]{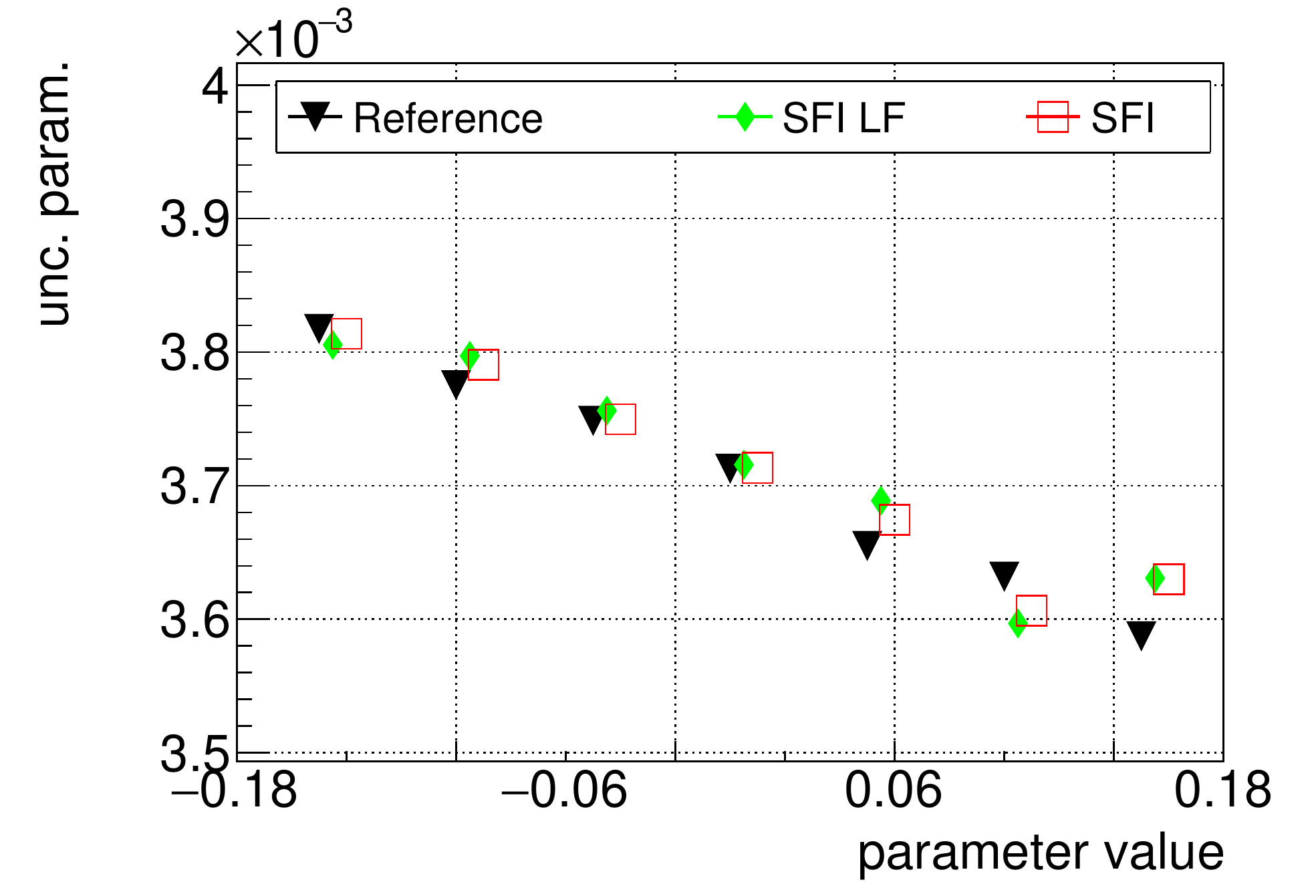}
  \includegraphics[width=0.45\textwidth]{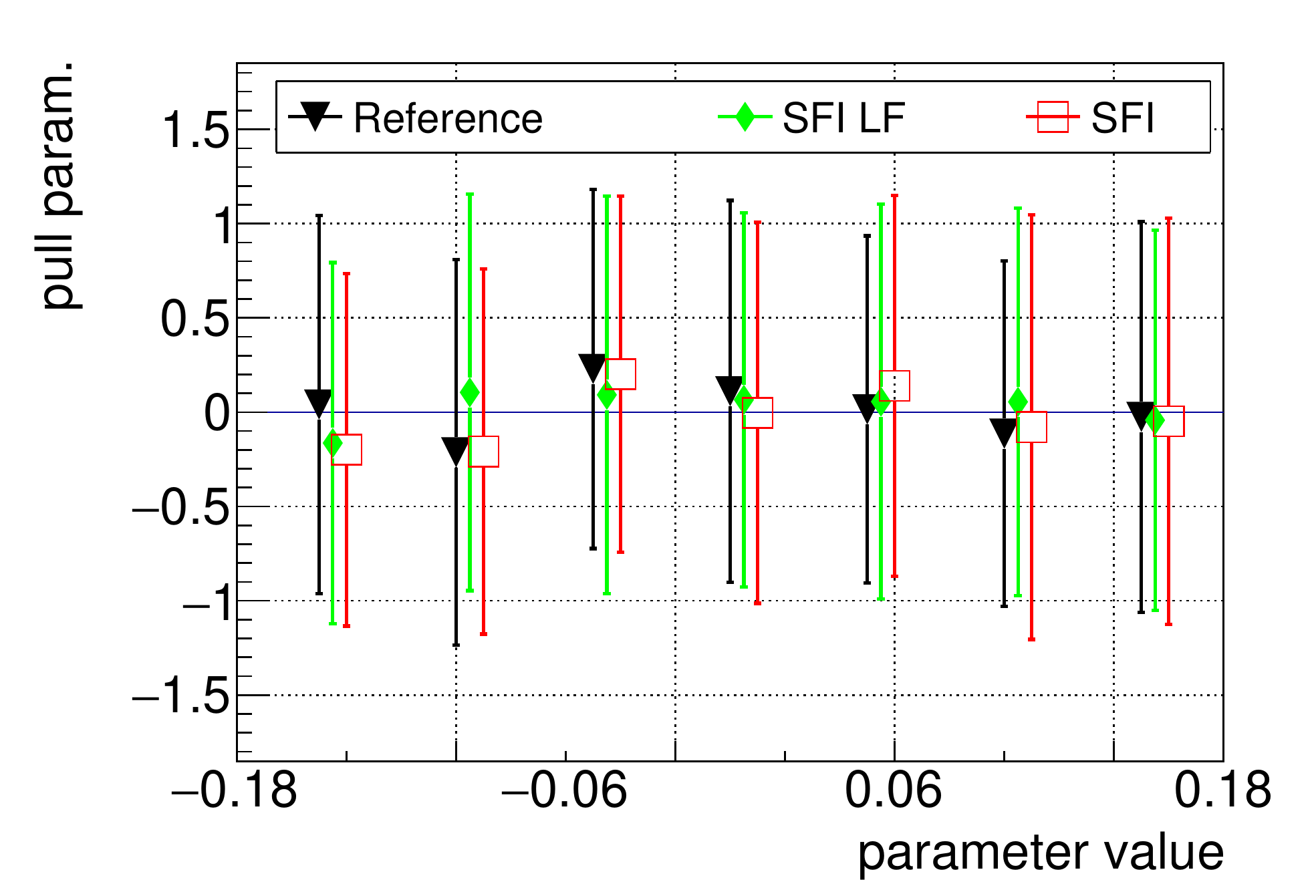}
  \caption{Uncertainty (left) and pull (right) for
    the parameter estimation example, comparing SFI to a reference provided
    by an unbinned fit of
    the original known function. 1k events are used for both signal and background. 
  }
  \label{fig:param_1D}
\end{figure}

\begin{figure}[!hbp]
  \centering
  \includegraphics[width=0.45\textwidth]{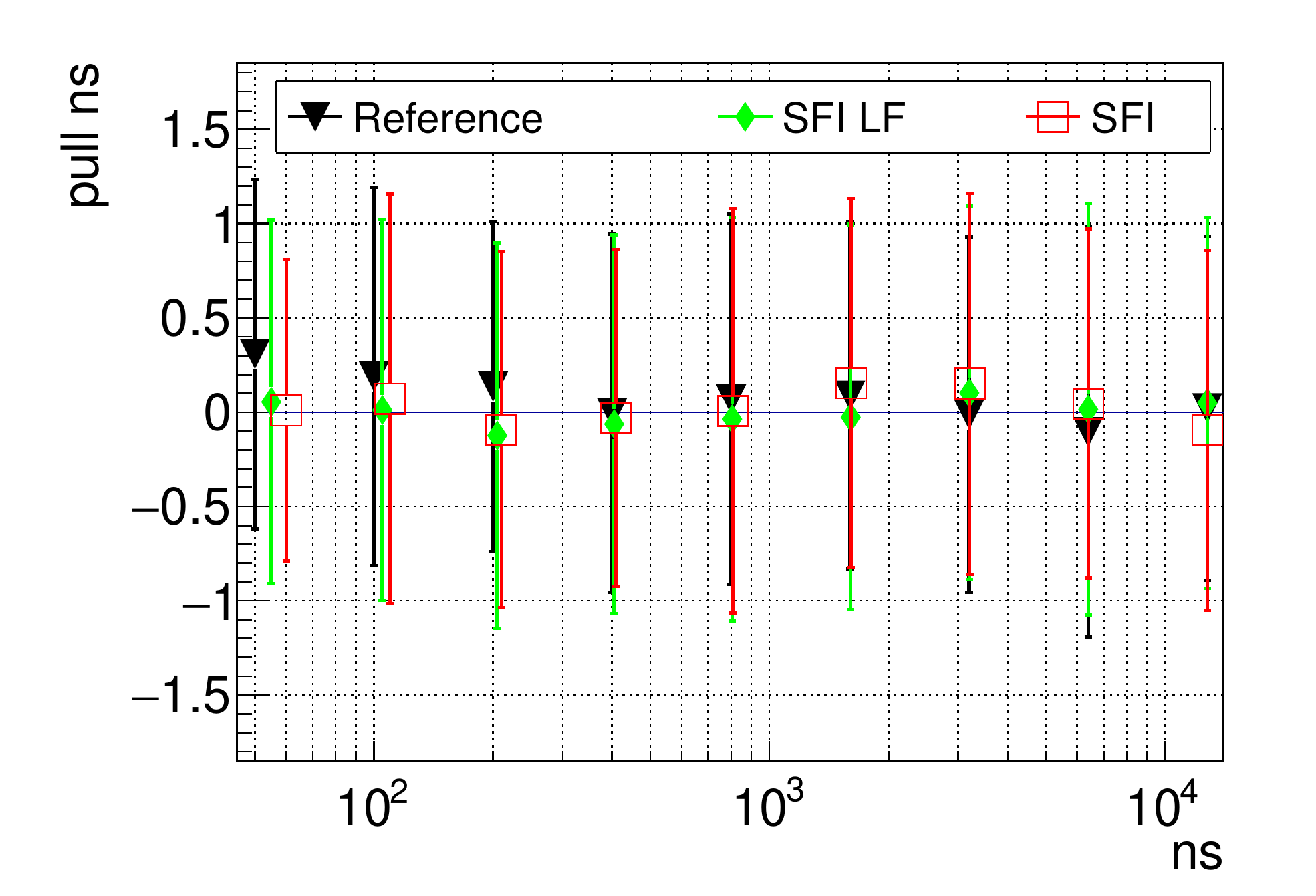}
  \includegraphics[width=0.45\textwidth]{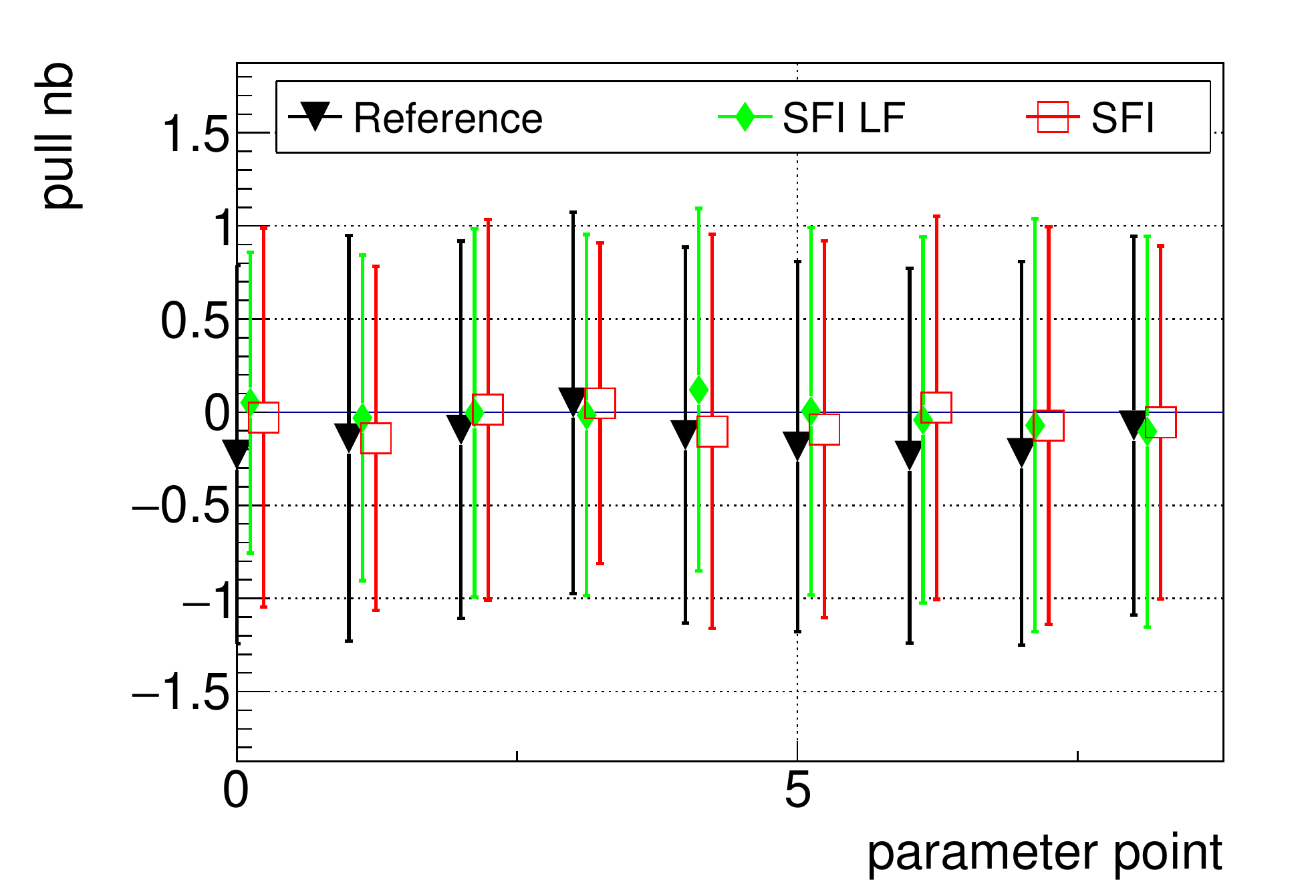}
  \caption{Pull for $\hat{n}_s$ (left) and $\hat{n}_b$ (right) in pseudo experiments where the values of $n_s$ have been varied, while $n_b$ was fixed. The value of the unknown parameter has been kept fixed.}
  \label{fig:param_1D_pull_N}
\end{figure}

The standard deviation and
pull from the parameter fit are shown in Figure \ref{fig:param_1D} and
compared to optimal performance. The parameter estimates are performed for 7 different values of $\textbf{c}$. Figure \ref{fig:param_1D_pull_N} shows pull for $\hat{n}_s$ and $\hat{n}_b$ in pseudo experiments where the number of signal events ($n_s$) have been varied while keeping a fixed number of background events ($n_b$). The true parameter value $\textbf{c}$ is constant in this case.

\subsubsection{When only discrete model point are available}\label{sec:discrete_c}
\begin{figure}[!hbp]
  \centering
  \includegraphics[width=0.45\textwidth]{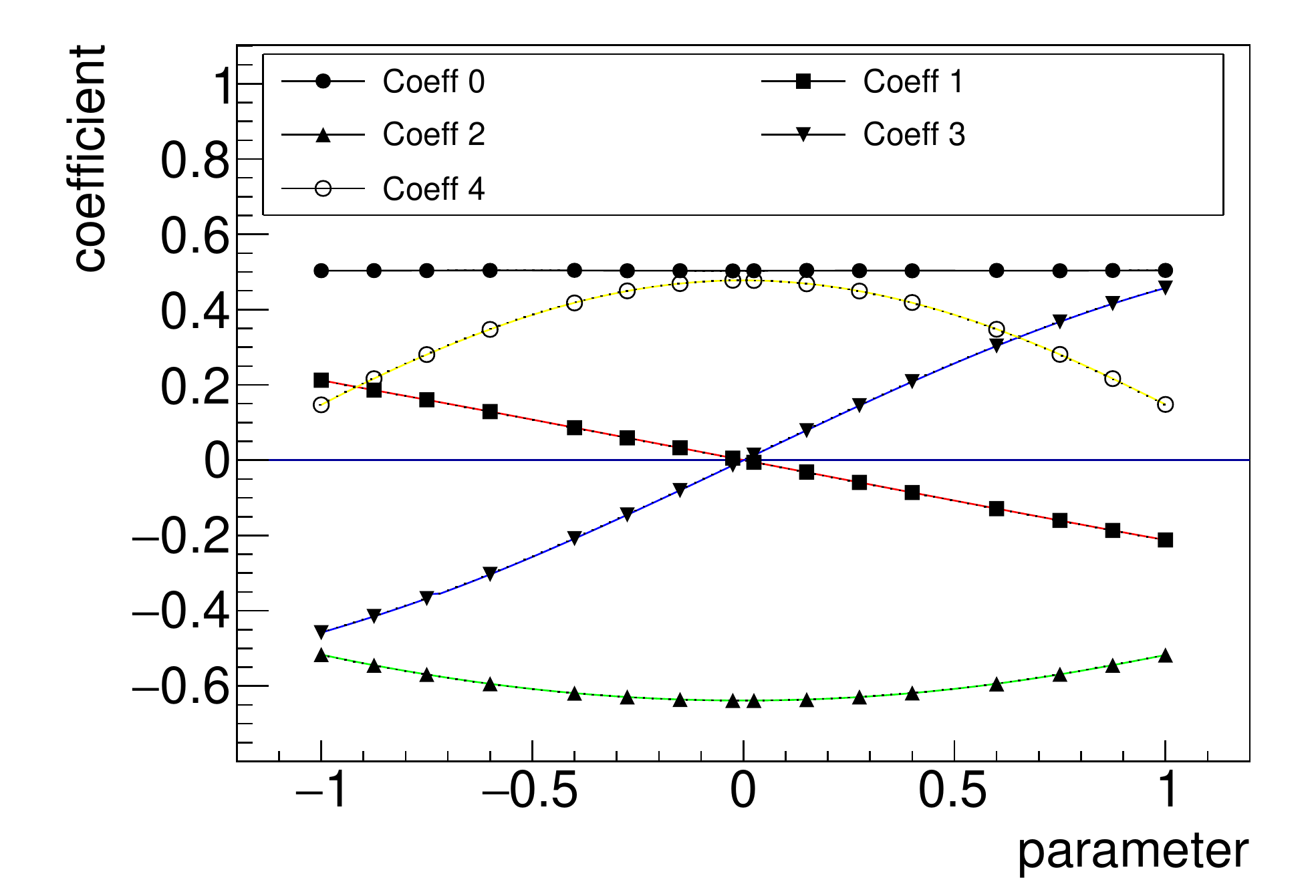}
  \caption{Coefficient interpolation for the square root transform as function of parameter value, in internal coordinates, between -1 and 1. Marked points are the coefficient values from the discrete model points while the line are the values interpolated from the combined transform derived with the technique outlined in Appendix \ref{fou}. }
  \label{fig:coefficient_interpolation}
\end{figure}

When the model is only available at certain grid points, i.e. discrete values of $\mathbf{c}$, which is common
when the model is given by Monte Carlo simulations, an interpolation
can be performed as follows.
First the Fourier coefficients are rewritten as a separate Fourier series
$b_i(\mathbf{c})=\sum_j c_{ij}\varphi_j(\mathbf{c})$.
The eigenfunctions $\varphi_j(\mathbf{c})$ are tensor products of one
dimensional Chebychev eigenfunctions. The cosine basis is used for
$\mathbf{x}$, $\phi_j(\mathbf{x})$,
and the full tensor product of $\varphi_j(\mathbf{c})$ and
$\phi_j(\mathbf{x})$ is used to build the series for
$p_s(\mathbf{x}_i, \mathbf{b}(\mathbf{c}))$.
Before data can be fitted, the Fourier coefficients $c_{ij}$ are first found by
SFI transforming each model sample at $\mathbf{c}_j$ and then solving a
linear equation
system, see Appendix \ref{fou} for details. This allows for interpolation
between discrete model points in case the model parameters cannot be generated
continuously. An interpolation example of the first coefficients in the
series is shown in Figure \ref{fig:coefficient_interpolation}. 
Note that this procedure can be applied to both the linear
and square root transforms.

\subsection{Linear SFI analysis in Fourier space}
In many cases it can be very advantageous to perform the analysis only within
Fourier space, i.e. use the Fourier coefficients instead of the function
values at the original $\textbf{x}_i$ coordinates. Unfortunately the
non-linear nature of the SFI square root transform
does not allow the coefficients to be directly compared between different
transformed distributions. Primarily due to different diagonalizations of the
covariance matrix and rescalings taking place for each transform. However,
in analogy with the characteristic function
which provides an alternative representation of the probability distribution
in Fourier space, the linear version of SFI (see Equation~\eqref{sfitransform})
allows for direct analysis in Fourier space.

The major drawback with linear SFI compared to default SFI is that the
coefficients are no longer uncorrelated.
The approach is based on the assumption that the linear SFI transformed
coefficients are approximately distributed as a
multivariate gaussian, which can be motivated by the central limit theorem.
The linear SFI likelihood\footnote{Note that this likelihood contains a factor -2} can then be written in Fourier space as
\begin{align}
  \label{linl}
l(n_s,n_b,\mathbf{c})=-2\ln(L(n_s,n_b,\mathbf{c}))=-\ln|\Sigma^{-1}(n_s,n_b,\mathbf{c})|+\mathbf{r}(n_s,n_b,\mathbf{c})^T\Sigma^{-1}(n_s,n_b,\mathbf{c})\mathbf{r}(n_s,n_b,\mathbf{c}),
\end{align}
where the residual $\mathbf{r}(n_s,n_b,\mathbf{c})=\mathbf{d}-n_s\mathbf{b}_s(\mathbf{c})-n_b\mathbf{b}_b$, and $\mathbf{d}$ is the vector with
filtered Fourier coefficients of the data. Both signal and background transforms are normalized to 1.
This formulation requires the \emph{precision matrix} $\Sigma^{-1}(n_s,n_b,\mathbf{c})$ to
be specified,
ideally analytically parameterized in the likelihood parameters $n_s$, $n_b$ and
$\mathbf{c}$.
Since $\Sigma^{-1}(n_s,n_b,\mathbf{c})$ a priori is unknown it has to be estimated
from the model, starting from an estimator $S(n_s,n_b,\mathbf{c})$ of the true
covariance matrix $\Sigma(n_s,n_b,\mathbf{c})$.
The estimated covariance matrix $S$ consists of two terms, a Poisson
distributed global normalization component $\Sigma_n$ for a fixed normalized shape
$$
\Sigma_n=n\mathbf{b}(\mathbf{c})\mathbf{b}^\text{T}(\mathbf{c}),
$$
and a noise component $\Sigma_S$ for fixed $n$ events with fluctuations
only due to shape variations. According to Equation \eqref{sfitransform} the one
event estimation of an eigenfunction is
$\phi_k(\mathbf{x}_i)=b_k^i$. The noise components $(\Sigma_S)_{kl}$ can be
measured using the sample covariance from each individual event as
$$
 (\Sigma_S)_{kl}=n \cdot cov(b_k, b_l) = \frac{n}{N}\sum_{\mathbf{x}_i \sim p(\mathbf{x})} (b^i_k - b_k)(b^i_l - b_l) = \frac{n}{N}\sum_{\mathbf{x}_i \sim p(\mathbf{x})} b^i_k b^i_l - n b_k b_l.
$$
$$
= \frac{n}{N}\sum_{\mathbf{x}_i \sim p(\mathbf{x})} \phi_k(\mathbf{x}_i) \phi_l(\mathbf{x}_i) - n b_k b_l
$$
This implies that $S$ can be written as
$$
S_{kl}=(\Sigma_n)_{kl}+(\Sigma_S)_{kl}=
\frac{n}{N}\sum_{\mathbf{x}_i \sim p(\mathbf{x})} \phi_k(\mathbf{x}_i) \phi_l(\mathbf{x}_i) \simeq n \left<\phi_k(\mathbf{x}) \phi_l(\mathbf{x})\right>_{p(\mathbf{x})}.
$$
The computational complexity of $S$ can further be drastically reduced by
using the product rule of orthogonal eigenfunctions which are of the form
$$
\phi_k \phi_l = \sum_{m(k,l)} g_m \phi_m,
$$
for some set of constants $g$. This allows for $S$ to be computed from the
already modelled coefficients $\mathbf{b}(\mathbf{c})$ as
$$
S_{kl}
=\frac{n}{N}\sum_{\mathbf{x}_i \sim p(\mathbf{x})} \sum_{m(k,l)} g_m\phi_m(\mathbf{x}_i) 
=n \sum_{m(k,l)} g_m b_m(\mathbf{c}).
$$
The covariance is composed of the covariances of the signal and background as:
$$
S(n_s, n_b,\mathbf{c}) = (n_s+n_b) \left(\frac{n_s}{n_s+n_b}\tilde{S}_s(\mathbf{c}) + \frac{n_b}{n_s+n_b}\tilde{S}_b
\right) = (n_s+n_b) \tilde{S}(n_s, n_b,\mathbf{c})$$
where $\tilde{S}_s(\mathbf{c})$ and $\tilde{S}_b$ are computed from the normalized coefficients. The likelihood can now be written:
$$
l(n_s,n_b,\mathbf{c})=\mathrm{dim}(\tilde{\Sigma}^{-1}(n_s, n_b,\mathbf{c}))\ln(n_s + n_b)-\ln|\tilde{\Sigma}^{-1}(n_s, n_b,\mathbf{c})|+\frac{1}{n_s + n_b}\mathbf{r}(n_s, n_b,\mathbf{c})^T\tilde{\Sigma}^{-1}(\mathbf{c})\mathbf{r}(n_s, n_b,\mathbf{c}).
$$

The precision matrix $\Sigma^{-1}(\mathbf{c})$ is then robustly reconstructed using regularization provided by the ROPE \cite{rope} recipe, controlled by a tuning parameter $\alpha$. The estimated covariance matrix $S$ is eigendecomposed and the eigenvalues of $\Sigma^{-1}$ are then computed (from the eigenvalues $\lambda^S_i$ of $S$) as:
$$\lambda_i = \frac{2}{\lambda^S_i + \sqrt{(\lambda^S_i)^2 + 8\alpha}} $$
unless $\lambda^S_i < 0.001 \times 8 \alpha$, in which case $\lambda_i = 0$. The regularized $\Sigma^{-1}$ corresponds to the maximum of the likelihood
$$
l_\mathrm{reg}(\Sigma^{-1}(n_s, n_b,\mathbf{c}),\alpha)=-\ln|\Sigma^{-1}(n_s, n_b,\mathbf{c})|+
\Tr(S\Sigma^{-1}(n_s, n_b,\mathbf{c}))+\alpha||\Sigma^{-1}(n_s, n_b,\mathbf{c})||^2_\mathrm{F}
$$
for all simulated events with a penalty term controlled by the
regularization parameter $\alpha$. The value of $\alpha$ is chosen to be $0.00001$.

Finally the parameters are extracted by minimizing the likelihood
$$
\arg\min_{n_s, n_b, \mathbf{c}} l(n_s, n_b, \mathbf{c}),
$$
where the precision matrix is calculated in each step during the minimization. The results for this method (SFI LF) are shown in Figures \ref{fig:param_1D} and \ref{fig:param_1D_pull_N}.

\section{Code implementation}
A C++ library and examples for some of the SFI calculations
explained in the text are available at the
\href{https://gitlab.com/sfi-lib/libsfi}{https://gitlab.com/sfi-lib/libsfi} repository. This library contains several sets of eigenfunctions, e.g. the Fourier basis can be used for periodic variables, and spherical harmonics for a pair of variables possessing spherical symmetry. Functionality for choosing eigenfunction set, as well as order, for each dimension is available. Together with variable transformations (such as arctan or the logarithm) and methods for reducing overtraining this allows the transform to be tailored to the general features of a multidimensional distribution.

In addition, a transform can be integrated to obtain the marginal density for a single dimension. The marginal density can then be compared to an empirical distribution, either by eye or by using a statistical test.

\section{Conclusions}
A general and efficient method for analysis of events in higher dimensions
using orthogonal real valued functions series is explained and exemplified.
The range of applications is large
and is not only limited to density estimation, classification
or parameter estimation.
In data analysis the reduction of degrees of freedom plays a
central role. A powerful and important key property of SFI is that for both the
square root and the linear transform the high degrees of freedom of the
original scattered data is reduced and encapsulated solely by the SFI Fourier
coefficients, no additional information is needed to specify the full covariance
matrix among the coefficients.

The range of applications falls into two distinct classes:
either the analysis is performed in the original $x$-space and only makes
a visit in Fourier space and comes back again (see the classification example),
or it leaves x-space completely and performs the analysis entirely in the
linearly transformed Fourier space (see the second parameter
estimation example). Note that the parameter
correlations present in the linear transform case is a restriction due to
increased complexity which is completely absent in the square root mode.
However, low complexity can still be maintained in the linear case by carefully
selecting the most relevant part of the spectrum from the signal point
of view.

A C++ code library
\href{https://gitlab.com/sfi-lib/libsfi}{https://gitlab.com/sfi-lib/libsfi} is provided with examples that allows for the reader to further explore other applications.

\section*{Acknowledgement}
This work was funded in part by the Knut and Alice Wallenberg foundation under grant no. KAW 2017.0100.

\bibliographystyle{unsrt}
\bibliography{\jobname.bib}

\appendix
\appendixpage

\section{Solving the model Fourier coefficients}
\label{fou}

The model, assumed to be dependent on the parameter vector $\mathbf{c}$, is
given at a discrete number of model points $\mathbf{c}_n$ and approximated as a finite
Fourier series as
$$
f(\mathbf{x},\mathbf{c}_n)=\sum_{ij}f_{ij}\phi_i(x)\varphi_j(\mathbf{c}_n),
$$
where $\varphi_j(\mathbf{c})$ here are chosen as Chebychev polynomials to
maximize the approximation efficiency. The basis set can either be SFI
or full tensor depending on the application.
The model points can also be SFI transformed individually without $\mathbf{c}$ dependence
as
$$
m^{n}(\mathbf{x})=\sum_k m^n_k \phi_k(\mathbf{x}).
$$
For a fixed model point $\mathbf{c}_n$ both descriptions should agree such that
(to reduce the notation $\varphi_j(\mathbf{c}_n)=\varphi^n_j$)
$$
\sum_k m^n_k \phi_k(\mathbf{x}) = \sum_{ij}f_{ij}\phi_i(\mathbf{x})\varphi^n_j.
$$
This constraint can be turned into a equation system in Fourier space by
projecting both sides on to the eigenfunction $\phi_l(\mathbf{x})$
$$
\int \sum_k m^n_k \phi_k(\mathbf{x}) \phi_l(\mathbf{x}) d\mathbf{x}= \int \sum_{ij}f_{ij}\phi_i(\mathbf{x})
\varphi^n_j \phi_l(\mathbf{x}) d\mathbf{x},
$$
which then becomes an equation system for the unknowns $f_{ij}$
$$
m^n_l =\sum_{j}f_{lj} \varphi^n_j.
$$
For $M=3$ Fourier dimensions in $\mathbf{x}$ and $N=2$ Fourier dimensions in
$\mathbf{c}$ the components of the equation system are
\begin{align}
m^0_0 &= \varphi^0_0 f_{00} + \varphi^0_1 f_{01} \nonumber \\
m^0_1 &= \varphi^0_0 f_{10} + \varphi^0_1 f_{11} \nonumber \\
m^0_2 &= \varphi^0_0 f_{20} + \varphi^0_1 f_{21} \nonumber \\
m^1_0 &= \varphi^1_0 f_{00} + \varphi^1_1 f_{01} \nonumber \\
m^1_1 &= \varphi^1_0 f_{10} + \varphi^1_1 f_{11} \nonumber \\
m^1_2 &= \varphi^1_0 f_{20} + \varphi^1_1 f_{21} \nonumber.
\end{align}
In matrix form this reads
$$
\begin{pmatrix}
m^0_0 \\
m^1_0 \\
m^0_1 \\
m^1_1 \\
m^0_2 \\
m^1_2
\end{pmatrix}
=
\begin{pmatrix}
\varphi^0_0 & \varphi^0_1 & 0 & 0 & 0 & 0 \\
\varphi^1_0 & \varphi^1_1 & 0 & 0 & 0 & 0 \\
0 & 0 & \varphi^0_0 & \varphi^0_1 & 0 & 0 \\
0 & 0 & \varphi^1_0 & \varphi^1_1 & 0 & 0 \\
0 & 0 & 0 & 0 & \varphi^0_0 & \varphi^0_1 \\
0 & 0 & 0 & 0 & \varphi^1_0 & \varphi^1_1
\end{pmatrix}
\begin{pmatrix}
f_{00} \\
f_{01} \\
f_{10} \\
f_{11} \\
f_{20} \\
f_{21}
\end{pmatrix}.
$$
From the structure of the components it is clear that the solution
consists of $M$ linear independent matrix equations, each of
size $N\times N$. To solve the system at least $N$ model points are needed.
In practice one should use more than $N$ points and use the least squares
solution to project the data down on to the truncated Chebychev series.

\section{Classification comparisons}
\label{class_cmp}
Tables \ref{spiral_efficiency} and \ref{hep_efficiency} contain comparisons between signal efficiencies for the different classification methods. The signal efficiencies are computed at certain background efficiencies (0.01, 0.1 and 0.3), both for the test and training samples. The goal is not only to get a high signal efficiency, but also to get compatible efficiencies in the test and training samples, i.e. to reduce overtraining. The aim has been to make a fair comparison, in terms of speed and efficiency, but it might be possible to further optimize the MLP and BDT methods to achieve the same -- or better -- performance in less time. Classification is however, as stated before, not the primary goal of the SFI methods.

\begin{table}[!hbp]
  \center
\begin{tabular}{l  r @{.} l  r @{.} l  r @{.} l  r @{.} l  r @{.} l  r @{.} l }
\hline
  & \multicolumn{2}{c}{Test @0.01}  & \multicolumn{2}{c}{Train @0.01}  & \multicolumn{2}{c}{Test @0.10}  & \multicolumn{2}{c}{Train @0.10}  & \multicolumn{2}{c}{Test @0.30}  & \multicolumn{2}{c}{Train @0.30} \\
\hline
SFI 
 & 0&71  & 0&72  & 0&96  & 0&96  & 0&99  & 0&99 \\
MLP 
 & 0&55  & 0&63  & 0&95  & 0&95  & 0&99  & 0&99 \\
BDT 
 & 0&62  & 0&67  & 0&95  & 0&95  & 0&99  & 0&99 \\
LD 
 & 0&13  & 0&14  & 0&20  & 0&20  & 0&57  & 0&54 \\
LDD 
 & 0&13  & 0&14  & 0&20  & 0&20  & 0&57  & 0&54 \\
\hline
\end{tabular}
\caption{Spiral signal efficiency for different background efficiencies\label{spiral_efficiency}}
\end{table}

\begin{table}[!hbp]
  \center
\begin{tabular}{l  r @{.} l  r @{.} l  r @{.} l  r @{.} l  r @{.} l  r @{.} l }
\hline
  & \multicolumn{2}{c}{Test @0.01}  & \multicolumn{2}{c}{Train @0.01}  & \multicolumn{2}{c}{Test @0.10}  & \multicolumn{2}{c}{Train @0.10}  & \multicolumn{2}{c}{Test @0.30}  & \multicolumn{2}{c}{Train @0.30} \\
\hline
SFI 
 & 0&44  & 0&45  & 0&92  & 0&93  & 0&99  & 0&99 \\
MLP 
 & 0&52  & 0&54  & 0&91  & 0&91  & 0&98  & 0&98 \\
BDT 
 & 0&57  & 0&66  & 0&92  & 0&93  & 0&98  & 0&99 \\
LD 
 & 0&17  & 0&16  & 0&60  & 0&61  & 0&89  & 0&91 \\
LDD 
 & 0&17  & 0&16  & 0&60  & 0&61  & 0&89  & 0&91 \\
\hline
\end{tabular}
\caption{HEP signal efficiency for different background efficiencies\label{hep_efficiency}}
\end{table}

\section{Configuration for the 2D spiral example}\label{apx:spiral_details}
The spirals are modelled by a 2-dimensional cosine transform with maximum degree 12, i.e. a monomic sparse tensor series. 10k events are used for both signal and background spiral, and half of the events are used for training and the other half for testing.

\section{Configuration for the HEP example}\label{apx:hep_details}
Both signal and background are modelled by a 6-dimensional transform of maximum degree 5. Overtraining is limited by performing significance pruning at a level of 3, i.e. coefficients where the value divided by the uncertainty is less than 3.0 are set to 0. 10k events are used for both signal and background samples, and half of the events are used for training and the other half for testing. The samples have been generated with MadGraph\cite{Alwall:2014hca}. The lepton transverse momentum $p_\texttt{T}$ is smeared by a gauss with width:
$$ \sqrt{(0.00038 p_\texttt{T})^2 + 0.015^2}.$$
Jet $p_\texttt{T}$ is smeared by a gauss with width:
$$ \sqrt{(0.5/\sqrt{E})^2 +0.03^2}.$$

The variables are constructed from two leptons (with $\eta < 2.5$) and the leading jet ($\eta < 2.5$) in the events. In collider physics it is common to use four dimensional vectors for particle energy and momentum, with components: E, $p_\texttt{T}$, $\eta$ and $\varphi$, where the momentum part is given as cylindrical coordinates with the z-axis along the axis of the colliding beams. The transverse momentum ($p_\texttt{T}$) is the component of the momentum perpendicular (or transverse) to the z-axis. The pseudorapidity ($\eta$) is, for approximately mass less particles, defined as $\eta = -\ln \tan \theta/2$, where $\theta$ is the angle between the particle momentum and the z-axis. $\varphi$ is the direction of the particle momentum in the x-y-plane.

The first variables are the lepton $p_\texttt{T}$ that are required to be in the range 20 to 100 GeV and they are log transformed. The third variable is the leading jet $p_\texttt{T}$ which is required to be in the range 20 to 200 GeV and is log transformed. The fourth variable is the invariant mass of the leptons, required to be in the range 21 to 161 GeV, and it is arctan transformed. The fifth and sixth variables are the lepton $\eta$ that are arctan transformed. For SFI the variable transformation types are the same but the parameters are different for the signal and background samples.

\section{Fourier and spherical harmonics bases}
Two sets of complex basis functions will be treated: The Fourier basis
\begin{equation}\label{eq:f_basis}
F_l(\phi) = n_l e^{i l \pi \phi}
\end{equation}
and spherical harmonics
\begin{equation}\label{eq:ylm_basis}
 Y^m_l(x, \phi) = n^m_{l} P^m_l(x) e^{i m \pi \phi}.
\end{equation} 
With variables $-1 \le \phi \le 1$ and $-1 \le x \le 1$. $n_l$ and $n^m_{l}$ are normalizations factors and they will be discussed below. $P^m_l$ are associated Legendre polynomials (ALP:s) that include the Condon-Shortley phase $(-1)^m$. Both basis functions have the property that $\phi$ is periodic. 

In most cases the variables will be omitted from the basis functions, $F_l, Y^m_l, \varphi_l$, unless special attention is needed. $\varphi_l$ will be used to denote a generic orthonormal basis function. Complex conjugation will be denoted $\bar{c}$. Unless specified, the indices in the series are summed as full tensors from $0$ to $N$, where $N$ is the maximum order of the basis functions, or as monomic sparse series.

Internally, in the SFI library, variables are transformed to the interval $(-1,1)$, either by a linear transformation or by one of the non-linear transformations. For the complex exponential the integral becomes:
\begin{equation}\label{eq:cplx_expo} 
\int_{-1}^{1} e^{im \pi \phi} d\phi = 2 \delta_{m0}.
\end{equation}

\subsection{Marginalizations}
In some situations it is desirable to perform marginalizations of a pdf of several variables. Besides statistical applications, e.g. a Kolmogorov-Smirnov test to data, it allows for visualization of the distributions and comparison to data by eye. 

For a linear transform, the marginal pdf is expressed using the integral ($I$) of the eigenfunctions:
$$ p(x_1, \mathbf{b}) = \sum_l b_l \varphi_l(x_1) = \int p(\mathbf{x},\mathbf{a}) dx_2 \ldots d x_d $$
$$ = \int \sum_{l_1 \ldots l_d}  a_{l_1 \ldots l_d} \varphi_{l_1}(x_1) \ldots \varphi_{l_d}(x_d) dx_2 \ldots d x_d  = \sum_{l_1 \ldots l_d}  a_{l_1 \ldots l_d} \varphi_{l_1}(x_1) I_{l_2} \ldots I_{l_d} $$

For the square root transform, since the eigenfunctions are required to be orthonormal, the marginalization procedure is simplified:
$$ p(x_1, \mathbf{b}) = \int p(\mathbf{x},\mathbf{a}) dx_2 \ldots d x_d  = \int A(\mathbf{x},\mathbf{a})^2 dx_2 \ldots d x_d$$
$$ = \sum_{l_1 \ldots l_d} \sum_{k_1 \ldots k_d} a_{l_1\ldots l_d} a_{k_1\ldots k_d} \int \varphi_{l_1}(x_1) \ldots \varphi_{l_d}(x_d) \varphi_{k_1}(x_1) \ldots \varphi_{k_d}(x_d) dx_2 \ldots d x_d $$
$$ = \sum_{l_1 \ldots l_d} \sum_{k_1} a_{l_1\ldots l_d} a_{k_1 l_2 \ldots l_d} \varphi_{l_1}(x_1) \varphi_{k_1}(x_1) $$

The next step is to turn the product $ \varphi_{l} \varphi_{k}$ into a sum of single $\varphi_l$. Such relations exist for Legendre polynomials and cosine, as well as for Fourier and spherical harmonics.

The marginalization procedure will thus turn a sqrt transform of several variables into a linear transform of fewer variables. Currently 1D and 2D marginalizations are supported, i.e. all but one or two variables may be integrated out.

\subsection{Fourier}
The implementation uses complex Fourier eigenfunctions. This means that
both coefficients and eigenfunctions are complex in the transform:
$$ p(x) \simeq \sum_{l=-N}^N a_l F_l. $$
Note that the index runs from negative to positive. 
However, since the transformed functions
for both the linear and the square root transform
are strictly real, there exists a reality constraint on the coefficients:
$ a_{-l} = \bar{a}_{l} $. Hence the the
expansion for a linear transform can be written as:
$$ p(x, \textbf{a}) = a_0F_0 \ldots + a_l F_l + \bar{a}_l \bar{F}_l + \ldots = a_0 F_0 + \sum_l 2 \texttt{re}\{a_l F_l\}.$$
This implies that the coefficients for negative $l$ are not needed (but it does \emph{not} imply that the coefficients are real.)
$A$ can also be written in terms of only positive $l$:
$$ A(x,\mathbf{b}) = b_0 \varphi_0 + 2 \texttt{re}\left\{ \sum_{l=1}^N b_l F_l \right\}.$$

To simplify the expressions, the eigenfunctions are scaled according to:
$$ \tilde{\varphi}_l(x) = c_l \varphi_l(x) \qquad c_l = \sqrt{2 - \delta_{l0}}. $$
A new set of coefficients for both the linear and the sqrt transform is obtained:
$$ \tilde{b}_l = c_l b_l .$$
Expressed in the scaled coefficients and eigenfunctions the sqrt transform becomes:
\begin{equation}\label{eq:amp_complex}
 A(x,\mathbf{b}) = \texttt{re}\left\{ \sum_{l} 2 d_l b_l F_l \right\} = \texttt{re}\left\{ \sum_{l} \tilde{b}_l \tilde{F}_l \right\}.
 \end{equation}
where $d_l$ has been introduced:
$$d_l = \frac{1}{1 + \delta_{l0}}.$$
with the property $2 d_l = c_l^2$.

The normalization factor $n_l$ is given by the integral \eqref{eq:cplx_expo} to be $1/\sqrt{2}$, and the scaled normalization becomes
$$\tilde{n}_l = c_l n_l = \frac{1}{\sqrt{1 + \delta_{l0}}}.$$

\subsubsection{Marginalization}
To properly account for coefficients with negative indices, rewrite \eqref{eq:amp_complex} in terms of the unscaled coefficients and eigenfunctions and take the square:
$$ A(x,\mathbf{b})^2 = \sum_{lk} d_l d_k ( b_l F_l + b_{-l} F_{-l})( b_k F_k + b_{-k} F_{-k}).$$
Introduce $ g_{lk} = b_l b_k F_l F_k = g_{kl}$ with properties $\bar{g}_{lk} = g_{-l-k}$  and $\bar{g}_{-lk} = g_{l-k}$:
$$ A(x,\mathbf{b})^2 = \sum_{lk} d_l d_k ( g_{lk} + g_{l-k} + g_{-lk} + g_{-l-k}) = \sum_{lk} 2 d_l d_k \texttt{re}\{ g_{lk} + g_{l-k} \}.$$

By using that $g_{lk} = g_{kl}$ and introducing a symmetry factor $e_{lk} = 2 - \delta_{lk}$, the series can be rewritten:
$$ A(x,\mathbf{b})^2 = \texttt{re}\left\{ \sum_{lk}^{k \le l} 2 d_l d_k  e_{lk} ( g_{lk} + g_{l-k}) \right\}.$$
The gain of this expression is that by exploiting symmetries, fewer operations have to be performed.

It now remains to reintroduce scaled coefficients and eigenfunctions. Products of Fourier eigenfunctions can be expressed as single eigenfunctions:
$$ F_l F_k = n_l n_k e^{i(l+k)\pi \phi} = \frac{n_l n_k}{n_{l+k}} F_{l+k} .$$
Which gives that:
$$ A(x,\mathbf{b})^2 = \texttt{re}\left\{ \sum_{lk}^{k \le l} 2 d_l d_k  e_{lk} \left( b_l b_k  \frac{n_l n_k}{n_{l+k}} F_{l+k} + b_l b_{-k}  \frac{n_l n_k}{n_{l-k}} F_{l-k} \right) \right\}.$$
With scaled coefficients and eigenfunctions:
$$ A(x,\mathbf{b})^2 = \texttt{re}\left\{ \sum_{lk}^{k \le l} \frac{2 d_l d_k  e_{lk} \tilde{n}_l \tilde{n}_k}{c_l^2 c_k^2} \left( \frac{\tilde{b}_l \tilde{b}_k}{\tilde{n}_{l+k}} \tilde{F}_{l+k} +   \frac{\tilde{b}_l \tilde{\bar{b}}_{k}}{\tilde{n}_{l-k}} \tilde{F}_{l-k} \right) \right\} $$
$$ = \texttt{re}\left\{ \sum_{lk}^{k \le l} \frac{e_{lk} \tilde{n}_l \tilde{n}_k}{2} \left( \frac{\tilde{b}_l \tilde{b}_k}{\tilde{n}_{l+k}} \tilde{F}_{l+k} +   \frac{\tilde{b}_l \tilde{\bar{b}}_{k}}{\tilde{n}_{l-k}} \tilde{F}_{l-k} \right) \right\}.$$
The advantage of this last expression is that the coefficients for the linear transform all have positive indices and are given in terms of the scaled coefficients. 

\subsection{Spherical harmonics}
The usual definition of $Y^m_l(x, \phi)$ is in terms of the angle $\theta$ such that $x = \cos(\theta)$, but the derivations will be clearer if this association is left out. The proper transformation will be achieved by making a variable transformation using cosine. 

Spherical harmonics \eqref{eq:ylm_basis} are defined for negative $m$, and those functions are related to the conjugate functions:
\begin{equation}\label{eq:ylm_conj}
\bar{Y}^m_l = (-1)^mY^{-m}_l.
\end{equation}
The linear transform can be written:
$$  p(x,\phi, \textbf{a}) = \sum_{lm}^{-l \le m \le l} a_{lm} Y^m_l. $$
We have that:
$$ a_{l-m} = (-1)^m \bar{a}_{lm} $$
and the coefficients for negative $m$ can be omitted, since:
\begin{IEEEeqnarray}{rCl}\label{eq:ylm_lin} 
p(x,\phi, \textbf{a}) & = & \sum_{lm}^{0 \le m \le l} d_m (a_{lm} Y^m_l +a_{l-m} Y^{-m}_l ) \nonumber \\
& = &\sum_{lm}^{0 \le m \le l} d_m (a_{lm} Y^m_l +\bar{a}_{lm} \bar{Y}^{m}_l ) \nonumber \\
 &= & \texttt{re}\left\{ \sum_{lm}^{0 \le m \le l} 2  d_m a_{lm} Y^m_l \right\} = \texttt{re}\left\{ \sum_{lm}^{0 \le m \le l} \tilde{a}_{lm} \tilde{Y}^m_l \right\}
\end{IEEEeqnarray}
Scaled coefficients and basis functions were introduced in the last step.

The normalization of $Y^m_l$ is given by:
$$ \int Y^m_l \bar{Y}^m_l d\Omega = \int_{-1}^{1} Y^m_l \bar{Y}^m_l dx d\phi = \int_{-1}^{1} (n^m_l)^2 P^m_l P^m_l dx d\phi = \frac{2(n^m_l)^2 2(l+m)!}{(2l+1)(l-m)!}=1.$$
$$ n^m_l = \sqrt{\frac{(2l+1)(l-m)!}{4(l+m)!}} $$
This differs from the usual normalization by a factor of $1/\sqrt{\pi}$ due to the different domain for $\phi$. With scaling, the normalization becomes:
$$ \tilde{n}^m_l = c_m n^m_l.$$

\subsubsection{Marginalization of a linear transform}
Expanding \eqref{eq:ylm_lin} in terms of the components of $Y^m_l$ \eqref{eq:ylm_basis} gives:
$$ p(x,\phi, \textbf{a}) = \texttt{re}\left\{ \sum_{lm}^{0 \le m \le l} c_m^2 a_{lm} n^m_{l}P^m_l e^{im\pi \phi} \right\}. $$
The marginalization in $x$ is given by:
$$ p(x, \textbf{b}) = \int_{-1}^{1} p(x,\phi, \textbf{a}) d\phi = \texttt{re}\left\{ \sum_{lm}^{0 \le m \le l} c_m^2 a_{lm} n^m_{l}P^m_l 2 \delta_{m0} \right\} = \texttt{re}\left\{ \sum_{l} 2 a_{l0} n^0_{l}P^0_l \right\}.$$
This is a series in Legendre polynomials($P_l$), where the coefficients and eigenfunctions are real:
$$ p(x,\textbf{b})= \sum_{l} b_l P_l = \sum_{l} \frac{2 a_{l0} n^0_{l}}{n^P_l} P_l  \qquad b_l = \frac{2 a_{l0} n^0_{l}}{n^P_l}$$
where $n^P_l$ are the normalization factors for Legendre polynomials.

The marginalization in $\phi$ involves integrals of ALP:s:
$$ p(\phi, \textbf{b}) = \int_{-1}^{1} p(x,\phi, \textbf{a}) dx = \texttt{re}\left\{ \sum_{lm}^{0 \le m \le l} c_m^2 a_{lm} n^m_{l} e^{im\pi \phi} \int_{-1}^{1} P^m_l dx \right\}. $$
The ALP integral is computed as \cite{DONG2002541}:
\begin{equation}\label{eq:alp_integral}
\int_{-1}^{1} P^m_l dx = \frac{((-1)^m + (-1)^l)2^{m-2}m\Gamma(l/2)\Gamma((l+m+1)/2)}{((l-m)/2)!\Gamma((l+3)/2)} = I^m_l .
\end{equation}
Reverting to scaled coefficients:
$$ p(\phi, \textbf{b}) = \texttt{re}\left\{ \sum_{lm}^{0 \le m \le l} \frac{\tilde{a}_{lm} \tilde{n}^m_{l} I^m_l}{\tilde{n}_m} \tilde{F}_m \right\} .$$
This shows that the marginalization results in a linear Fourier series.

\subsubsection{Marginalization of a sqrt transform}
The sqrt transform in spherical harmonics is written as:
$$ A(x,\phi,\textbf{a})^2 = \sum_{lmkn} d_m d_n (a_{lm} Y^m_l +\bar{a}_{lm} \bar{Y}^{m}_l) (a_{kn} Y^n_k +\bar{a}_{kn} \bar{Y}^{n}_k) .$$
Introducing $g_{lmkn} = a_{lm} a_{kn} Y^m_l Y^n_k$ gives:
$$ A(x,\phi,\textbf{a})^2 = \sum_{lmkn} d_m d_n (g_{lmkn} + g_{l-mkn} + g_{lmk-n} + g_{l-mk-n}) .$$

Using the properties of the coefficients and basis functions we have $ g_{l-mkn} = \bar{g}_{lmk-n}$ and $g_{l-mk-n} = \bar{g}_{lmkn}$.
$$ A(x,\phi,\textbf{a})^2 = \sum_{lmkn} d_m d_n (g_{lmkn} + \bar{g}_{lmkn} + g_{lmk-n} + \bar{g}_{lmk-n}) = \texttt{re}\left\{\sum_{lmkn} 2 d_m d_n (g_{lmkn} +  g_{lmk-n}) \right\}.$$

If the serie is required to only have positive $m$, then one has to impose the condition $n \le m$ and introduce the symmetry factor $e_{mn} = 2 - \delta_{mn}$. Expanding $g_{lmkn}$ and using that $a_{k-n} = (-1)^n\bar{a}_{kn}$ gives:
\begin{IEEEeqnarray}{rCl}\label{eq:ylm_amp} 
A(x,\phi,\textbf{a})^2 &=& \texttt{re}\left\{\sum_{lmkn}^{n \le m} 2 e_{mn} d_m d_n ( a_{lm} a_{kn} Y^m_l Y^n_k +  (-1)^n a_{lm} \bar{a}_{kn} Y^m_l Y^{-n}_k) \right\}.
\end{IEEEeqnarray}

The relation \eqref{eq:ylm_amp} has three applications: Marginalization of a multidimensional transform into a linear 2D transform in $x$ and $\phi$ and into a 1D transform of either $x$ or $\phi$. 

For the first case, the product of two spherical harmonics can be turned into a sum of single spherical harmonics. Begin with the expression for a product of two ALP:s \cite{DONG2002541}:
\begin{equation}\label{eq:Plm_product}
P^m_l P^n_k = \sqrt{\frac{(l+m)!(k+n)!}{(l-m)!(k-n)!}}\sum^{l+k}_{L=max(M, |l-k|)} \sqrt{\frac{(L-M)!}{(L+M)!}} C^{l k L}_{000}C^{l k L}_{m n M} P^M_L.
\end{equation}
Where $M=m+n$, $C^{l k L}_{m n M}$ are Clebsch-Gordan coefficients, and the constraint $M \le L$ has been explicitly included. With this the expression for the product of the spherical harmonics can be written
\begin{IEEEeqnarray}{rCl}\label{eq:ylm_product}
 Y^{m}_{l}Y^{n}_{k} &=& \sqrt{\frac{(2l + 1)(2k + 1)}{16}}\sum_{L=max(M,|l - k|)}^{l+k} \sqrt{\frac{4}{(2L+1)}} C^{l k L}_{000}C^{l k L}_{m n M} Y^{M}_L \\
 &=& \sum_{L=max(M,|l - k|)}^{l+k} q_{lmknL}Y^{M}_L.
\end{IEEEeqnarray}
Note the difference in normalization. 

Using this in the expression for $A$ gives:
$$ A(x,\phi,\textbf{a})^2 $$
$$= \texttt{re}\left\{\sum_{lmkn}^{n \le m} 2 e_{mn} d_m d_n \left( a_{lm} a_{kn} \sum_L q_{lmknL} Y^{m+n}_L +  (-1)^n a_{lm} \bar{a}_{kn} \sum_L q_{lmk-nL} Y^{m-n}_L\right) \right\}$$
 $$= \texttt{re}\left\{\sum_{lmkn}^{n \le m} \frac{2 e_{mn} d_m d_n}{c_m c_n} \left( \tilde{a}_{lm} \tilde{a}_{kn} \sum_L \frac{q_{lmknL}}{c_{m+n}} \tilde{Y}^{m+n}_L +  (-1)^n \tilde{a}_{lm} \bar{\tilde{a}}_{kn} \sum_L \frac{q_{lmk-nL}}{c_{m-n}} \tilde{Y}^{m-n}_L\right)  \right\}
$$
where $L$ is summed from $max(M,|l - k|)$ to $l+k$.
This expression can be used to implement marginalization of a multi dimensional transform containing spherical harmonics.

Turning to the marginalization in $x$, expanding \eqref{eq:ylm_amp} in the parts of $Y^m_l$ gives:
$$
 \int A(x,\phi,\textbf{a})^2 d\phi =
$$
$$
 \texttt{re}\left\{\sum_{lmkn}^{n \le m} 2 e_{mn} d_m d_n n^m_l n^n_k \left( a_{lm} a_{kn} P^m_l P^n_k \int_{-1}^{1} e^{i(m+n)\pi\phi} d\phi +  a_{lm} \bar{a}_{kn} P^m_l P^{n}_k \int_{-1}^{1} e^{i(m-n)\pi\phi} d\phi \right) \right\} \nonumber
 $$

The integral of the complex exponentials are only nonzero for the first term if $m+n=0$ and $m-n=0$ for the second. Since both $m$ and $n$ are positive we have $2\delta_{mn}\delta_{m0}$ for the first term and $2\delta_{mn}$ for the second. Both have the condition that $m\le k$ in addition to $m \le l$. It then follows that:
\begin{equation}\label{eq:ylm_marg_x}
\int A(x,\phi,\textbf{a})^2 d\phi  = \texttt{re}\left\{\sum_{lmk}^{m \le k} 4 e_{mm} d_m d_m n^m_{l} n^m_{k} \left( a_{lm} a_{km} P^m_l P^m_k \delta_{m0} +  a_{lm} \bar{a}_{km} P^m_l P^{m}_k \right) \right\}
\end{equation}
$$ = \texttt{re}\left\{\sum_{lmk}^{m \le k} \frac{4 d_m d_m \tilde{n}^m_{l} \tilde{n}^m_{k}}{c_m^4} \left( \tilde{a}_{lm} \tilde{a}_{km} P^m_l P^m_k \delta_{m0} +  \tilde{a}_{lm} \bar{\tilde{a}}_{km} P^m_l P^{m}_k \right) \right\} $$
$$ = \texttt{re}\left\{\sum_{lmk}^{m \le k} \tilde{n}^m_{l} \tilde{n}^m_{k} \left( \tilde{a}_{lm} \tilde{a}_{km} P^m_l P^m_k \delta_{m0} +  \tilde{a}_{lm} \bar{\tilde{a}}_{km} P^m_l P^{m}_k \right) \right\} $$

Using  \eqref{eq:Plm_product}, the last expression can be turned into a linear transform of ALP:s. The use of this transform is limited however, since the ALP:s are \emph{not} orthonormal.

The marginalization in $\phi$ is given by a transform in the Fourier basis:
$$ \int A(x,\phi,\textbf{a})^2 dx = $$
$$\texttt{re}\left\{\sum_{lmkn}^{n \le m} 2 e_{mn} d_m d_n n^m_l n^n_k \left( a_{lm} a_{kn} e^{i(m+n)\pi\phi} \int P^m_l P^n_k dx+  a_{lm} \bar{a}_{kn} e^{i(m-n)\pi\phi} \int P^m_l P^{n}_k dx \right) \right\} $$
Using \eqref{eq:Plm_product} and \eqref{eq:alp_integral} the integral of two ALP:s ($I^{mn}_{lk}$) can be computed, which gives:
$$ \int A(x,\phi,\textbf{a})^2 dx = \texttt{re}\left\{\sum_{lmkn}^{n \le m} \frac{2 e_{mn} d_m d_n \tilde{n}^m_l \tilde{n}^n_k I^{mn}_{lk}}{c_m^2 c_n^2} \left( \frac{\tilde{a}_{lm} \tilde{a}_{kn}}{\tilde{n}_{m+n}} \tilde{F}_{m+n} +  \frac{\tilde{a}_{lm} \tilde{\bar{a}}_{kn}}{\tilde{n}_{m-n}} \tilde{F}_{m-n} \right) \right\} $$
$$ = \texttt{re}\left\{\sum_{lmkn}^{n \le m} \frac{e_{mn} \tilde{n}^m_l \tilde{n}^n_k I^{mn}_{lk}}{2} \left( \frac{\tilde{a}_{lm} \tilde{a}_{kn}}{\tilde{n}_{m+n}} \tilde{F}_{m+n} +  \frac{\tilde{a}_{lm} \tilde{\bar{a}}_{kn}}{\tilde{n}_{m-n}} \tilde{F}_{m-n} \right) \right\} $$

\subsection{Computations}
In \eqref{eq:ylm_basis} there are three quantities that need to be computed. The fourier factor is computed using:
$$ F_n = 2 \cos(\pi\phi) F_{n-1} - F_{n-2}$$
with $F_0 = 1$ and $F_1 = e^{i\pi\phi}$.

The ALP is computed using the following, which can be found in \cite{NIST:DLMF} or derived from relations found there:
$$ P^{l-1}_l = x (2l - 1) P^{l-1}_{l-1} $$
$$ P^l_l = - \sqrt{1 - x^2} (2l - 1) P^{l-1}_{l-1} $$
$$ P^m_l = \frac{-1}{(l+m+1)(l-m)} \left( 2\frac{(m+1)x}{\sqrt{1 - x^2}} P^{m+1}_l + P^{m+2}_l \right) $$
with the initial values:
$$ P^0_0 = 1 \qquad P^0_1 = x \qquad P^1_1 = - \sqrt{1 - x^2}. $$
There are alternative recursion relations, but the one chosen was found to be more accurate than a recursion starting with $m=0$.

 The Clebsh-Gordan coefficients are computed in a dense table. There are symmetry properties:
 $$ C^{l k L}_{m n M} = (-1)^{l+k-L} C^{k l L}_{n m M} = (-1)^{l+k-L} C^{k l L}_{(-n) (-m) (-M}) $$
 from which we can determine $k \le l$ and $0 \le M$, i.e. we do not need to compute coefficients with $M < 0$ or $k > l$. Further there are constraints:
 $$ -l \le m \le l \qquad -k \le n \le k $$
 $$ l-k \le L \le l+k \qquad 0 \le M = m + n \le L$$
From the last constraint it is clear that $n$ (or $m$) can be omitted. Note that $m$ and $n$ are allowed to take on negative values. Using $n = M - m$ there are two constraints on $m$: $ -k \le M -m \le k$ and $-l \le m \le l$. The first one can be written as $M-k\le m \le M+k$. Examining the lower bound reveals that $-l \le M-k$ since $k\le l$ and $0\le M$. The number of coefficients can be found from:
$$ N_c = \sum_{l=0}^N\sum_{k=0}^{k \le l} \sum_{L=l-k}^{l+k} \sum_{M=0}^{L} \sum_{m=M-k}^{\min(l,M+k)} 1$$
where $N$ is the maximum number for $l$. An index in the table may be computed from this:
$$ I(l,k,L,M,m) = 
\sum_{l'=0}^{l-1}\sum_{k'=0}^{k' \le l'} \sum_{L'=l'-k'}^{l'+k'} \sum_{M'=0}^{L'} \sum_{m'=M'-k'}^{\min(l',M'+k')} 1 
+ \sum_{k'=0}^{k-1} \sum_{L'=l'-k'}^{l'+k'} \sum_{M'=0}^{L'} \sum_{m'=M'-k'}^{\min(l',M'+k')} 1 $$
$$ + \sum_{L'=l'-k'}^{L-1} \sum_{M'=0}^{L'} \sum_{m'=M'-k'}^{\min(l',M'+k')} 1 + \sum_{M'=0}^{M-1} \sum_{m'=M'-k'}^{\min(l',M'+k')} 1 + \sum_{m'=M'-k'}^{m-1} 1
$$  
The actual computations of the coefficients are done using recursion relations from \cite{Racah:104181}:
\begin{equation}\label{eq:cg_rec}
C_{\pm}(L,M) C^{l k L}_{m n (M\pm 1)} = C_{\pm}(l, m\mp1) C^{l k L}_{(m\mp 1) n M} + C_{\pm}(k, n \mp 1) C^{l k L}_{m (n\mp 1) M}
\end{equation}
with:
$$ C_{\pm}(l,m) = \sqrt{l(l+1) - m(m \pm 1)}.$$
The initial recursion relation can be found by using \eqref{eq:cg_rec} with $M=L$ and $C_+$:
$$ 0 = C_+(l, m-1) C^{l k L}_{(m-1) (n+1) L} + C_+(k, n) C^{l k L}_{m n L} \quad \leftrightarrow $$
\begin{equation}\label{eq:cg_initial}
C^{l k L}_{(m-1) (n+1) L} = -\frac{C_+(k, n)}{C_+(l, m-1)} C^{l k L}_{m n L} = D_{(m-1),n}C^{l k L}_{m n L}
\end{equation}
Here we have $L-k \le m \le l$ and $n = L - m$. The initial constraint is that:
$$ \sum_{m=L-k}^{l} (C^{l k L}_{m n L})^2 = 1 = ((\ldots+1)D^2_{(l-1),(L-l)} + 1)(C^{l k L}_{l (L-l) L})^2$$
from which we get $C^{l k L}_{l (L-l) L}$ and by repeatedly using \eqref{eq:cg_initial} the rest of the coefficients for $M=L$ can be computed. To get the coefficients for $M < L$ use \eqref{eq:cg_rec} with $C_{-}$.
\end{document}